\documentclass{article} %
\usepackage[nonatbib,preprint]{neurips_2025}

\usepackage{times}

\usepackage{url}

\usepackage{CJKutf8}

\usepackage{booktabs}       %
\usepackage{amsfonts}       %
\usepackage{nicefrac}       %
\usepackage{microtype}      %
\usepackage{mathtools}
\usepackage{multirow}
\usepackage{bm}
\usepackage{epsfig}
\usepackage{graphicx}
\usepackage{subcaption}
\usepackage{amsthm}
\usepackage{amsmath}
\usepackage{amssymb}
\usepackage{enumitem}
\usepackage{makecell}
\usepackage{wrapfig}
\usepackage{indentfirst}
\usepackage{verbatim}
\usepackage{color}
\usepackage{setspace}
\usepackage{array}
\usepackage{stackengine}

\usepackage{algorithm}
\usepackage[algo2e,algoruled,boxed,lined]{algorithm2e}
\usepackage{algorithmicx}
\usepackage{algcompatible}

\usepackage{graphicx}
\usepackage{xcolor}
\usepackage{anyfontsize}

\usepackage{caption}

\usepackage[pagebackref=true,breaklinks=true,colorlinks=true,bookmarks=false]{hyperref}
\definecolor{deepred}{HTML}{940000}
\hypersetup{linkcolor=deepred}
\hypersetup{urlcolor  = [rgb]{0.4,0.15,0.95}}
\hypersetup{citecolor=[rgb]{0.4,0.15,0.95}}

\definecolor{model}{HTML}{00639E}
\definecolor{opt}{HTML}{2A5F1C}

\usepackage{colortbl}
\definecolor{Gray}{gray}{0.94}
\newcolumntype{a}{>{\columncolor{Gray}}c}
\newcolumntype{Y}{>{\centering\arraybackslash}X}

\newlength\savewidth

\usepackage{tabularx}

\usepackage{pifont}
\usepackage{tocloft}
\usepackage[toc,page,header]{appendix}
\usepackage{adjustbox}
\usepackage{minitoc}

\renewcommand \thepart{}
\renewcommand \partname{}

\usepackage[capitalize,noabbrev]{cleveref}

\usepackage{lipsum}

\usepackage{epigraph}

\usepackage{listings}

\title{Automating RT Planning at Scale: High Quality Data For AI Training}
\author{
  Riqiang Gao\textsuperscript{1}  \quad Mamadou Diallo\textsuperscript{1} \quad Han Liu\textsuperscript{1} \quad Anthony Magliari\textsuperscript{2} \quad Jonathan Sackett\textsuperscript{2}\\
  \bf
  \quad Wilko Verbakel\textsuperscript{2}  
 \quad Sandra Meyers\textsuperscript{3} \quad  Rafe Mcbeth\textsuperscript{4} \quad Masoud Zarepisheh\textsuperscript{5}  \\
  \bf
  \quad Simon Arberet\textsuperscript{1} \quad Martin Kraus\textsuperscript{1} \quad 
  Florin C. Ghesu\textsuperscript{1} \quad Ali Kamen\textsuperscript{1}\\[1.5mm]
  \textsuperscript{1} Digital Technology and Innovation, Siemens Healthineers \\
  \textsuperscript{2}Varian Medical Systems, Siemens Healthineers\\
  \textsuperscript{3}UCSD,
  \textsuperscript{4}UPenn,
  \textsuperscript{5}MSKCC\\
}

\begin{document}
\maketitle
\doparttoc 
\faketableofcontents
\vspace{-0.1in}
\begin{abstract}

Radiotherapy (RT) planning is complex, subjective, and time-intensive. Advances with artificial intelligence (AI) promise to improve its precision and efficiency, but progress is often limited by the scarcity of large, standardized datasets. To address this, we introduce the Automated Iterative RT Planning (AIRTP) system, a scalable solution for generating high-quality treatment plans. This scalable solution is designed to generate substantial volumes of consistently high-quality treatment plans, overcoming a key obstacle in the advancement of AI-driven RT planning. 
Our AIRTP pipeline adheres to clinical guidelines and automates essential steps, including organ-at-risk (OAR) contouring, helper structure creation, beam setup, optimization, and plan quality improvement, using AI integrated with RT planning software like Varian’s Eclipse. Furthermore, a novel approach for determining optimization parameters to reproduce 3D dose distributions, i.e. a method to convert dose predictions to deliverable treatment plans constrained by machine limitations is proposed. A comparative analysis of plan quality reveals that our automated pipeline produces treatment plans of quality comparable to those generated manually, which traditionally require several hours of labor per plan. Committed to public research, the first data release of our AIRTP pipeline includes nine cohorts covering head-and-neck and lung cancer sites to support an AAPM 2025 challenge.  To our best knowledge, this dataset features \textbf{ more than 10 times} number of plans compared to the largest existing well-curated public dataset. 

\vspace{0.03in}
\href{https://huggingface.co/datasets/Jungle15/Radiotherapy_HaN_Lung_AIRTP}{\textit{DICOM Data}} \hspace{0.18in} 
\href{https://huggingface.co/datasets/Jungle15/GDP-HMM_Challenge}{\textit{Processed Data}} \hspace{0.18in} \href{https://github.com/RiqiangGao/GDP-HMM_AAPMChallenge}{\textit{GitHub}} \hspace{0.18in} \href{https://qtim-challenges.southcentralus.cloudapp.azure.com/competitions/38/}{\textit{Challenge}} \hspace{0.18in} \href{https://huggingface.co/Jungle15/GDP-HMM_baseline}{\textit{Pretrained Model}}  

\end{abstract}

\section{Introduction}
\label{sec:intro}
Radiation therapy (RT) is a critical treatment modality used in approximately 50\% of all cancer cases. Deep learning approaches have been incorporated into various stages of the RT planning process, including dose prediction \cite{kui2024review,Babier2020OpenKBP:Challenge,gao2023flexible,zhang2024dosediff}, fluence prediction \cite{Wang2020FluenceTherapy,arberet2025beam}, leaf sequencing \cite{gao2024multi,hrinivich2024clinical}, and dose calculation \cite{xing2020feasibility}. Among recent advancements in RT, 3D dose prediction emerges as a particularly promising application. This approach directly estimates 3D dose distributions from inputs such as CT scans, planning target volume (PTV)/organ at risk (OAR) masks, and planning configurations, eliminating the need for labor-intensive optimization processes. Dose prediction serves multiple purposes, including supporting optimization objectives \cite{Babier2020OpenKBP:Challenge,Babier2022OpenKBP-Opt}, enabling quality assurance \cite{Gronberg2023DeepPlans,Gronberg2023DeepCancers}, and functioning as a key component in AI-driven planning pipelines \cite{fan2019automatic,gao2024multi}. 

\begin{figure}
    \centering
    \includegraphics[width=0.95\linewidth]{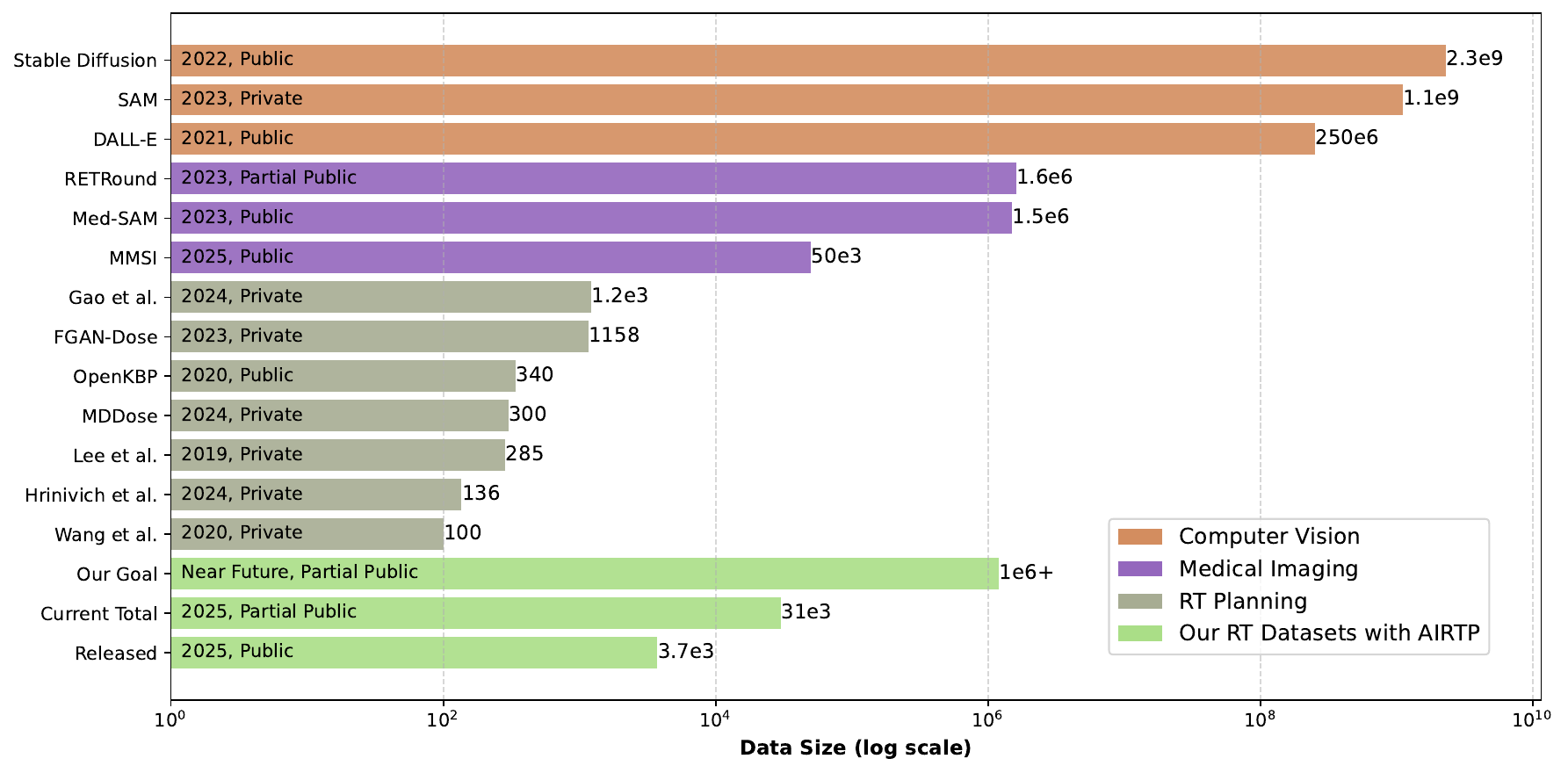}
    \caption{Datasets of representative AI models across various domains (see Appendix \ref{sec:data_scale} for details). Most existing works in ``RT Planning'' with AI are based on approximately 1,000 or fewer treatment plans, significantly smaller than the data scale used in representative successful AI models. Additionally, the majority of RT studies rely on private datasets, limiting reproducibility and scalability. ``Our RT Datasets with AIRTP'' aim to bridge data gaps between RT research and vision domains.}
    \label{fig:data_scale}
    \vspace{-0.2in} 
\end{figure}

Data curation of AI for RT planning faces significant challenges related to both data \textbf{\textit{quality}} and \textbf{\textit{quantity}}. \textbf{\textit{Quality}} limitations arise from several perspectives: (1) subject-level inconsistency, due to substantial variability in how individual clinicians contour PTVs / OARs and conduct treatment planning; (2) cohort-level inconsistency, as different cohorts may follow distinct clinical guidelines and institutional practices, introducing systematic biases (see Sec. \ref{sec:rtchallenge}); (3) protocol-level inconsistency, whereby plan quality fluctuates over time within the same institution due to upgrades in optimization technologies and changes in treatment protocols \cite{tol2016longitudinal}; and (4) the lack of standardized and quantitative metrics to objectively assess plan quality, further complicating dataset normalization. In addition, the difficulty of collecting RT data in clinical practice further exacerbates \textbf{\textit{quantity}} limitations. Unlike domains where large-scale datasets have fueled major advances in AI, RT planning datasets typically contain a relatively small number of patients, constraining model generalization and robustness (see Fig. \ref{fig:data_scale} and Appendix \ref{sec:data_scale}). Together, these quality and quantity challenges can present significant barriers to next advanced generation AI models for RT planning.


There are opportunities to streamline scalable high-quality planning with recent achievements.  AI-based organ segmentation has achieved significant advancements in both research and commercial applications \cite{azad2024medical}. Researchers and dosimetrists publish RapidPlan$^{\text{TM}}$ \cite{varian2024rapidplan} models to estimate planning objectives following clinical guidelines \cite{varian2024rapidmodels}. Varian Medical Systems’ Eclipse, a widely used platform for optimizing radiotherapy treatment plans for cancer patients, supports automation through its Eclipse Scripting API (ESAPI) \cite{varian2024esapi}. ESAPI enables developers to access and manipulate patient data for planning purposes, streamlining traditionally manual tasks performed via user interface. 

In this work, we present an \textbf{a}utomatic \textbf{i}terative \textbf{r}adio\textbf{t}herapy \textbf{p}lanning (AIRTP) pipeline to address RT data \textbf{\textit{quality}} and \textbf{\textit{quantity}} limitations for advanced AI modeling. 
The key idea of AIRTP pipeline is to automatically replicate the human RT planning process by effectively integrating existing tools and our proposed strategies. In brief, our main contributions are be summarized as follows: 

\begin{itemize}
\label{adv}
\item Our AIRTP pipeline dramatically cuts planning time from the typical 3–6 hours required manually to just 0.1–1 hour automatically, facilitating large-scale AI training with high-quality treatment plans. It is also generalizable across various treatment sites.
\item We introduce an novel iterative refinement strategy that aligns plan quality to customized scorecard without requiring human intervention.
\item We develop a method to replicate 3D dose distributions in different scenarios using AIRTP, facilitating evaluation of dose prediction impacts on deliverable RT planning. 
\item We have released the Heterogenous Multi-cohort and Multi-site Radiotherapy Planning (HMM-RT) dataset for an AAPM challenge\footnote{Our GDP-HMM challenge is one of \href{https://www.aapm.org/GrandChallenge}{top two proposals} selected and sponsored by the American Association of Physicists in Medicine (AAPM) Working Group on Grand Challenges.} and beyond. This dataset: 1) contains over 10 times more plans than the existing curated public dataset \cite{Babier2020OpenKBP:Challenge}, and 2) offers more practical settings to support diverse research explorations (see Table \ref{tab:vsOpenKBP}).
\end{itemize}

\section{Background and Related Work} 
\label{sec:rtchallenge}

\textbf{Radiotherapy Planning} is a complex and highly specialized process that requires close coordination among clinical experts, including radiation oncologists, medical physicists, and dosimetrists \cite{Huynh2020ArtificialOncology}. In practice, it typically begins with a simulation CT scan, followed by manual contouring of PTVs and OARs, prescription definition, configuration of planning parameters, initial plan generation, physician review, and iterative refinement. The planning configuration may include choosing of plan type: Intensity-Modulated Radiation Therapy (IMRT) versus Volumetric Modulated Arc Therapy (VMAT), and beam geometries.  The entire planning process usually involves 3–6 hours of active work over 1 to 5 days. Furthermore, it relies heavily on manual input and subjective clinical judgment, making systematic data collection particularly challenging.

\textbf{Challenges of Clinical Plans for AI Training}. The variability in RT planning data poses significant challenges for AI modeling training. This variability stems from differences in PTV/OAR contouring styles, planning preferences, auxiliary structures, clinical protocols, and human biases \cite{verbakel2019targeted}. 
\begin{table}[]
    \scriptsize
    \centering
    \renewcommand{\arraystretch}{1.5}
    \caption{Properties of OpenKBP \cite{Babier2020OpenKBP:Challenge} and our HMM-RT   
 datasets. The \#patients and \#plans are  numbers released in GDP-HMM challenge; total number of our datasets for internal research is larger. }
    \begin{tabular}{llll}
    \toprule
     Attribute  & OpenKBP \cite{Babier2020OpenKBP:Challenge} & Our HMM-RT & Comments \\
    \midrule
      \cellcolor{Gray}Treatment Site & \cellcolor{Gray}HaN & \cellcolor{Gray}HaN \& Lung & \cellcolor{Gray} \\
      Prescribed Doses & Fixed & Variable & \parbox{7cm}{OpenKBP change original clinical intent (PTV contours and prescription) to make data homogeneous.  We keep the original clinical intent with variable prescription dose.}  \\
      \cellcolor{Gray}Number of PTVs  & \cellcolor{Gray}Up to 3 & \cellcolor{Gray}Up to 3 &  \cellcolor{Gray} \\
      Planning Tool \
      [year] & CERR [2003] & Eclipse [v.2023] & \parbox{7cm}{Eclipse is mature product from a leading radiotherapy company (i.e.,Varian) and widely used in clinical practice.} \\
      \cellcolor{Gray}\# Plans & \cellcolor{Gray}340 & \cellcolor{Gray} 3730 &\cellcolor{Gray}  \\
      \# Patients & 340 & 1622 &  \\
      \cellcolor{Gray} Quality Assessment & \cellcolor{Gray}No & \cellcolor{Gray}Yes & \cellcolor{Gray}\parbox{7cm}{We use the quality score card to evaluate the plan quality and remove or re-plan corner cases, make plans more aligned to scorecard.} \\
      \# Structures & up to 10/site & up to 50/site & Structures in our pipeline are Table \ref{tab:contours} and \ref{tab:lungobj}. \\
      \cellcolor{Gray}Format/Type & \cellcolor{Gray}CT/Struct/Dose & \cellcolor{Gray}CT/Struct/Plan/Dose & \cellcolor{Gray}\parbox{7cm}{We provide both processed Numpy data and RAW DICOMs exported from Eclipse, which support various research topics; while OpenKBP provide data only in csv format and lost many clinical information for research topics beyond dose prediction.} \\
      
    \end{tabular}
    \label{tab:vsOpenKBP}
\end{table}
Differences in contouring—such as variations in margins, anatomical boundaries, or slice intervals—introduce noise and bias into the data, limiting a model’s ability to generalize across institutions and clinicians. Moreover, non-standardized planning preferences, including differences in dose constraints and optimization strategies, produce diverse and often conflicting data inputs. As illustrated in Fig.~\ref{fig:oar_cnt}, the number of clinically contoured OARs per patient varies considerably (from 2 to 24), complicating efforts to normalize plan quality to a consistent standard.

This lack of uniformity significantly hinders the development of robust AI models that can learn effectively across multiple clinical settings. Further compounding the issue, much of the available RT datasets miss explicit definitions of optimization objectives, making it difficult to align contoured structures with their corresponding RT dose distributions. Additionally, Table~\ref{tab:rtnames} highlights the heterogeneity in RT structure naming, which leads to noisy associations between inputs and outputs, further complicating model training.

Additionally, some clinical plans were created more than five or ten years ago, when radiotherapy technology was less advanced. As a result, these plans may not meet the higher quality standards enabled by current technology. This highlights needs for re-planning RT treatments using the latest advancements while preserving clinical intents.

\textbf{Related Work}. A closely related effort to our work is from OpenKBP challenge \cite{Babier2020OpenKBP:Challenge}. Table \ref{tab:vsOpenKBP} outlines the key differences between our approach and the OpenKBP data curation pipeline. Our work advances AI-based dose prediction by enhancing treatment planning across several dimensions. Specifically, we address more practical and diverse scenarios, including a broader range of treatment sites, variable prescribed doses tailored to individual patients, and an expanded set of ROI structures. Our dataset is also substantially larger, containing over ten times the number of plans compared to OpenKBP. Additionally, our plans are generated using one of the latest and most advanced planning systems—Eclipse 18.0.1 (released in 2023 by Varian Medical Systems). Unlike OpenKBP, our methodology aligns more with current clinical guidelines for both planning objectives and quality assessment, ensuring greater relevance to real-world clinical practice.

Knowledge-based planning (KBP) is also related to our AIRTP pipeline \cite{meyer2021automation}. Both traditional KBP methods (e.g., RapidPlan$^{\text{TM}}$) \cite{varian2024rapidplan,van2018personalized,hytonen2022influence} and deep-learning methods (e.g., \cite{mcintosh2017fully,fan2019automatic,Babier2020OpenKBP:Challenge, gao2023flexible}) leverage previous knowledge to inform current planning, automating conventional planning processes to some extent. However, these techniques have not yet been utilized for large-scale deliverable planning. For example, RapidPlan is integrated into the Eclipse system, but it is primarily used to estimate optimization objectives, and the overall planning process still requires significant manual effort. In contrast, our study incorporates a high-quality RapidPlan model \cite{magliari2024hn} as part of AIRTP, automating the entire planning process and enhancing the quality of RapidPlan through iterative refinement.

\section{Method: Automatic High Quality Planning at Scale}
\label{sec:method}
\begin{figure}
    \centering
    \includegraphics[width=0.97\linewidth]{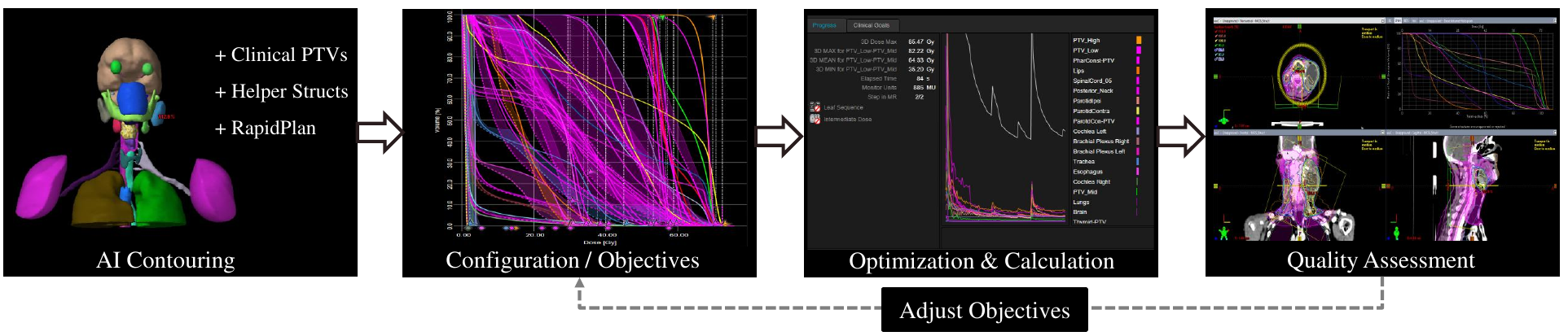}
    \caption{Illustration of automated iterative RT planning (AIRTP) pipeline. The auto contouring, helper structures are executed with C++/python code. Curated CT and RTSTRUCT data are stored in DICOM format, and then imported into Eclipse system. Beam configuration, RapidPlan setup, photon optimization, dose calculation, quality assessment are conducted with Eclipse Scripting API. A full data curation pipeline are appended in Fig. \ref{fig:curation}.}
    \label{fig:pipeline}
    \vspace{-0.1in}
\end{figure}

Fig. \ref{fig:pipeline} shows major steps of our AIRTP pipeline for scalable RT planning \footnote{An introduction of RT planning and terminologies can be found Appendix A of \cite{gao2024multi}.} and we introduce more details in the following. Full data curation pipeline with AIRTP integration is appended in Fig. \ref{fig:curation}.

\textbf{Step 1: RT Structures Creation.} As shown in Fig. \ref{fig:pipeline}, we first segment all the needed OARs 
based on model from AI Rad Companion of Siemens Healthineers \cite{siemens2024airad}. Following guidelines in \cite{magliari2024hn}, we create PTV helpers including ring structures and OAR$_{i}$-PTV, which are part of the RapidPlan \cite{magliari2024hn} optimization dose objective derivation. Structures and the associated descriptions are appended in Table \ref{tab:contours} and \ref{tab:lungobj}. Patients can have up to 52 structures and all structures are saved in DICOM format and then imported with CT series to the Eclipse workstation.   

\textbf{Step 2: Planning Scripting.} 
Four major modules are scripted using ESAPI. The first module, \textit{beam configuration}, involves setting up the number of fractions, beam geometries, and machine parameters. The second module defines \textit{planning objectives} based on a combination of the RapidPlan model, Scorecard metrics, and iterative refinement, which will be discussed in more detail later. For the \textit{plan optimization} and \textit{dose calculation}, we utilize the TrueBeam system with Photon Optimization version PO.18.0.1 and Dose Calculation AXB version 18.0.1, respectively. To balance computational efficiency and optimization quality, optimization convergence is set to ``On" for VMAT plans.

For head-and-neck cancer treatment, we use four arcs for VMAT, alternating gantry rotations (CW and CCW) and setting collimator positions at 30° and 330°. For IMRT, we create two plans per patient using 9 and 15 evenly spaced angles. The objective definitions of the initial plan are primarily based on the RapidPlan model, following structure mapping. The RapidPlan model is obtained from the official Varian Medical Affairs website \cite{varian2024han}. 

For lung cancer treatment, we use three different angle ranges based on the tumor's location, measured by the lateral distance from the isocenter to the patient's mid-sagittal plane. If the distance exceeds 5 cm, we select either the left or right lung template, depending on the tumor's side. If the distance is 5 cm or less, we choose angles that cover the full range. For IMRT, we use 7 fields (if lateral) or 9 angles, and for VMAT we use two arcs. Visual examples can be found in Fig. \ref{fig:example} and Appendix \ref{sec:supp_moreexp}.
 
\begin{algorithm}[tb]
\small
\caption{Iterative Planning Processing }\label{alg:iter}
\begin{algorithmic}
\STATE {\bfseries Input:} Patient data with CT, PTV/OAR contours, helper structures, prescribed doses for PTVs; ScoreCard for plan evaluation; beam configurations; RapidPlan for a head-and-neck scenario.  
\STATE {\bfseries Output:} A series of plans until the ScoreCard convergence. 
\end{algorithmic}
\begin{algorithmic}[1]
   \STATE Initialize plan based on RapidPlan or ScorePlan items 
   \FOR{$n$ = 1 to \textit{MAX iteration}}
   \STATE Initialize objective buffer $D$
    \FOR{$s$ in \textit{PTVs and its ring structures}}
    \STATE Set the upper / lower point objectives based prescribed dose
    \STATE Use the same priorities as in the initial plan  
    \ENDFOR
    \STATE Calculate the ScoreCard for current plan
   \FOR{$s$ in \textit{ScoreCard OARs and helper structures}}
   \STATE Calculate the DVH Cumulative Data $f_s(\cdot)$
    \FOR{$v$ in \textit{volume points \{1, 10, 20, 30, 40, 60, 80\}}}
    \STATE Calculate the dose $d = f_s(v)$
    \STATE Calculate the margin and objective dose and priority based on appended Fig. \ref{code:dose2obj}. 
    \STATE Add the objective to objective buffer.
    \ENDFOR 
    \FOR{items in \textit{ScoreCard}}
    \STATE Calculate the margin and objective dose and priority based on appended Fig. \ref{code:scorecard2dose}. 
    \STATE Add the objective to objective buffer.
    \ENDFOR 
   \ENDFOR
   \STATE Optimize with the current objective buffer. 
   \STATE Update current plan. 
 \ENDFOR
\end{algorithmic}
\end{algorithm}

 \begin{figure}
    \centering
    \begin{tabular}{c}
        \includegraphics[width = 0.97\textwidth]{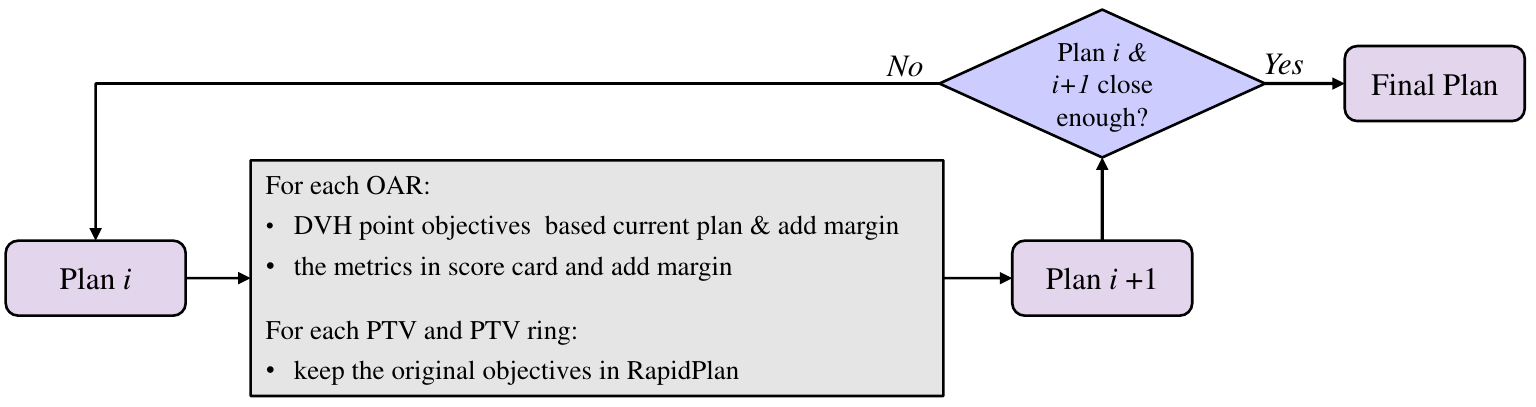} \\
        (a) Iterative high-quality planning. \\
        \includegraphics[width =0.97\textwidth]{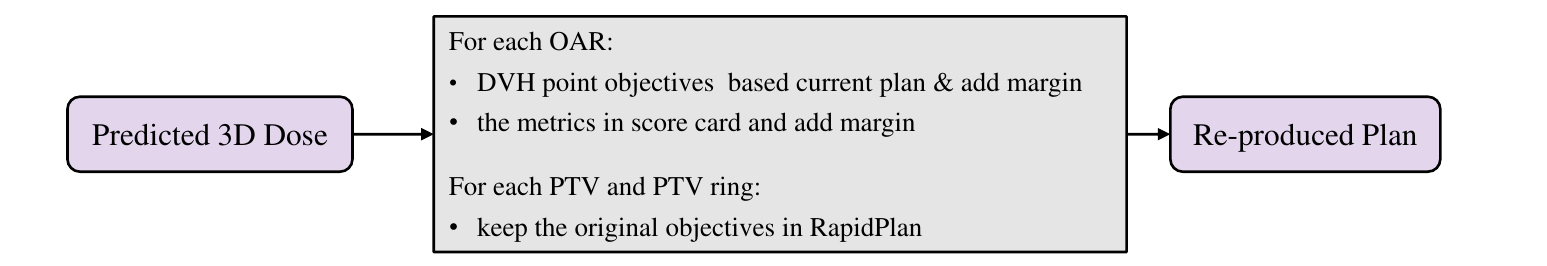}\\
        (b) Translate 3D dose prediction to deliverable plan. \\
    \end{tabular}
    \caption{(a) shows the process that increase the quality of planning iteratively. (b) demonstrates the method translating predicted 3D dose to a deliverable plan. The way of defining objectives of above two pipelines are the same. }
    \label{fig:iterativedose}
    \vspace{-0.1in}
\end{figure}

\textbf{Step 3: Iterative Process.} The initial treatment plan is generated using either RapidPlan \cite{magliari2024hn} for head-and-neck cases or by applying mean/point objectives derived from scorecard metrics \cite{varian2024lungconv} for lung cases, as detailed in Table \ref{tab:lungscard}. The iterative optimization workflow is illustrated in Fig. \ref{fig:iterativedose}(a). For each organ-at-risk (OAR), dose-volume histogram (DVH) points \cite{Drzymala1991DOSE-VOLUMEHISTOGRAMS} are extracted from the current plan, and margins are applied to update the optimization objectives. Scorecard metrics are also integrated into the objective functions with appropriate margins.

When the parameters are properly tuned, empirical evidence shows that plan quality tends to converge toward a high score. As illustrated in Fig. \ref{fig:iterativedose}(b), following the pipeline in Fig. \ref{fig:iterativedose}(a) allows us to reproduce RT plans using only the 3D dose distribution—an ability leveraged for evaluating dose prediction models (Sec. \ref{sec:reproduce}).

More detailed steps of this process are presented in Algorithm \ref{alg:iter}, with selected module implementations shown in Fig. \ref{code:dose2obj} and Fig. \ref{code:scorecard2dose}. This iterative framework is designed to mimic the clinical workflow, incorporating expert review and quality assurance steps. The AIRTP pipeline is modular and flexible, enabling easy integration of improved submodules (e.g., updated AI contouring models) into the overall system.

\section{Planning and Results}

\subsection{Planning Environment}

The first step in RT structure creation involves auto-contouring of OARs and the creation of helper structures, primarily based on clinical PTVs. The auto-contouring part utilizes the latest version of internal AI-Rad Companion \cite{siemens2024airad}. To accelerate this process, we developed a C++ executable that generates all necessary structures with a naming convention and exports the results in DICOM format.

To run auto-planning at scale, we have built five GPU computing nodes in Miscorsoft Azure, in which instances of Eclipse 18.0.1 are installed. Each node has 64 GB Random-access memory (RAM) and 24 GB GPU memory. We use PyESAPI \cite{varian2024PyESAPI} and Python to import DICOM input data and run the scripted algorithm to interact with the Eclipse optimization engine. PyESAPI is a Python-based tool that  enables rapid prototyping of C\# based ESAPI built-in functionalities. 

As described in Sec. \ref{sec:method} and illustrated in Fig. \ref{fig:curation}, we utilize CT images, PTVs, and prescribed dose information extracted from clinical data. The public TCIA datasets used in this study include NSCLC-Cetuximab \cite{bradley2018data}, NSCLC-Radiomics \cite{aerts2014data}, Head-Neck Cetuximab \cite{bosch2015head}, Head-Neck-PET-CT \cite{vallieres2017data}, HNC-IMRT-70-33 \cite{Buatti2024CT}, HNSCC \cite{grossberg2020hnscc}, HNSCC-3DCT-RT \cite{bejarano2018head}, and TCGA-HNSC \cite{zuley2016cancer}.

\begin{figure}
    \centering
    \includegraphics[width=0.95\linewidth]{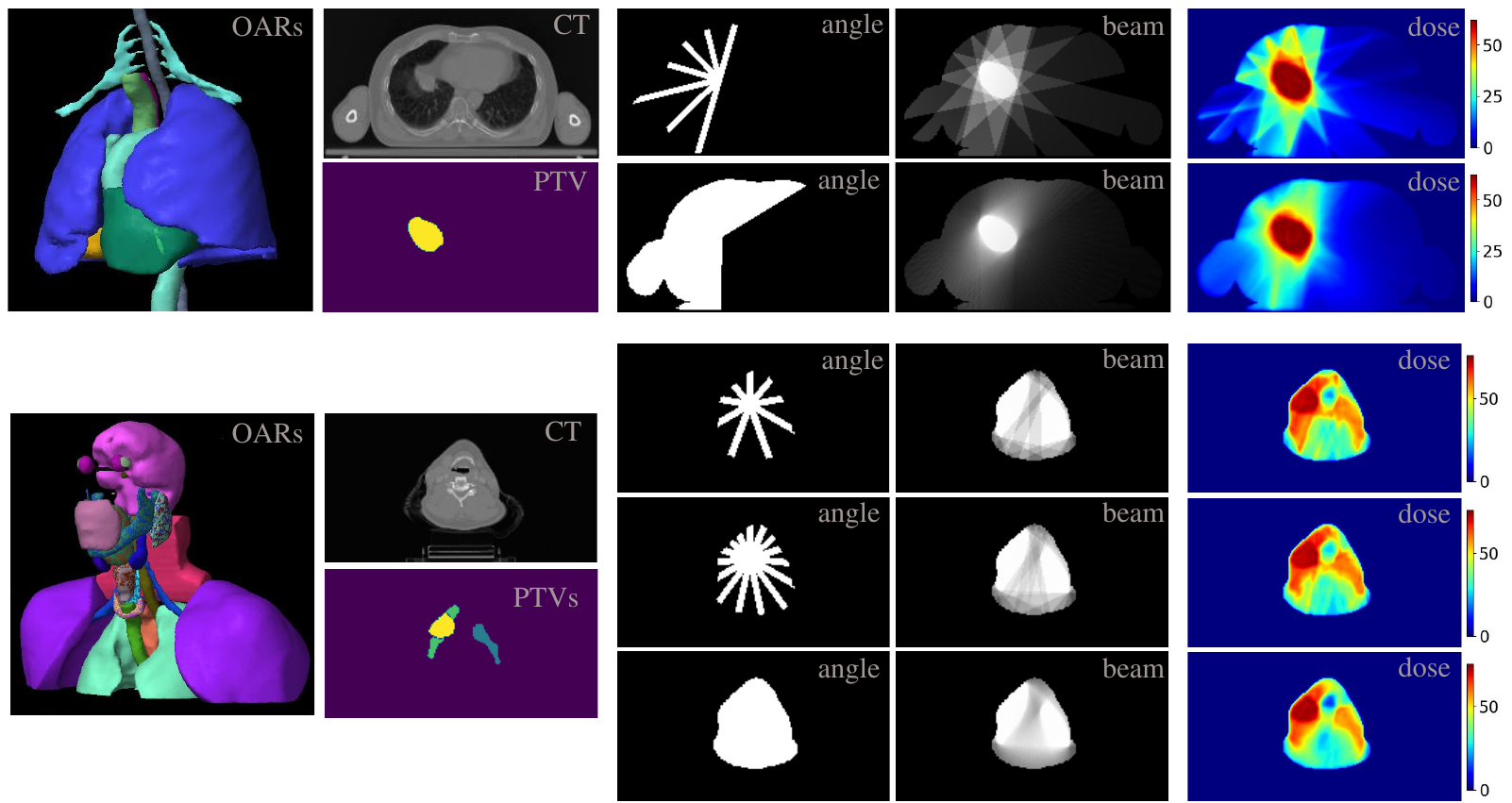}
    \caption{Examples of head-and-neck (upper) and lung (lower) cancer treatment plans. In released HMM-RT dataset, each lung cancer patient is assigned two plans: one IMRT and one VMAT; while each head-and-neck cancer patient receives three plans: two IMRT and one VMAT.}
    \vspace{-0.1in}
    \label{fig:example}
\end{figure}


\subsection{Plan Quality Discussion}

Fig. \ref{fig:example} shows plan examples for patients with lung cancer (Lung) and head and neck cancer (HaN), including CT images, structures, angle \& beam plates \cite{gao2023flexible}, and dose. Our AIRTP can generate arbitrarily large number of plans with varying parameters, given sufficient time and computing resources. As the first public release of our data, we include two plans per patient for Lung and three plans per patient for HaN. Additional examples can be found in Appendix \ref{sec:supp_plate} and \ref{sec:supp_moreexp}.


\begin{figure}
\centering
\begin{subfigure}{0.49\textwidth}
    \includegraphics[width=\textwidth]{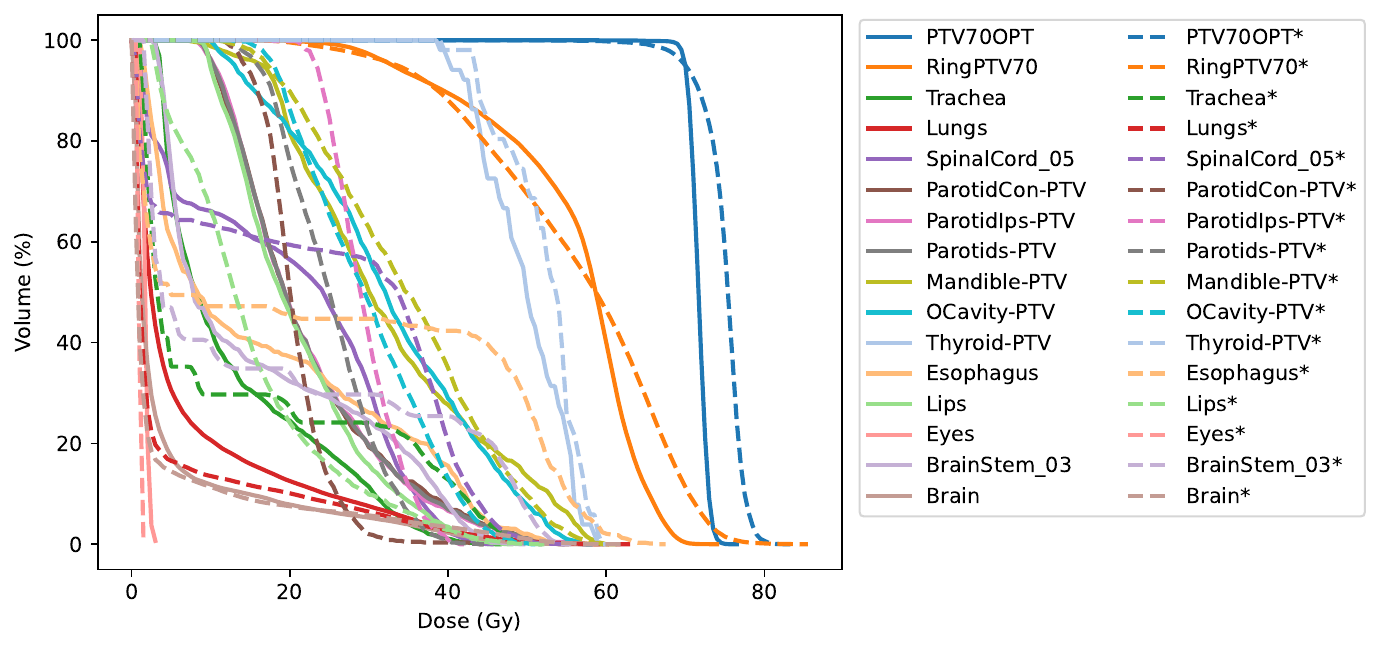}
    \label{fig:first}
\end{subfigure}
\hfill
\begin{subfigure}{0.49\textwidth}
    \includegraphics[width=\textwidth]{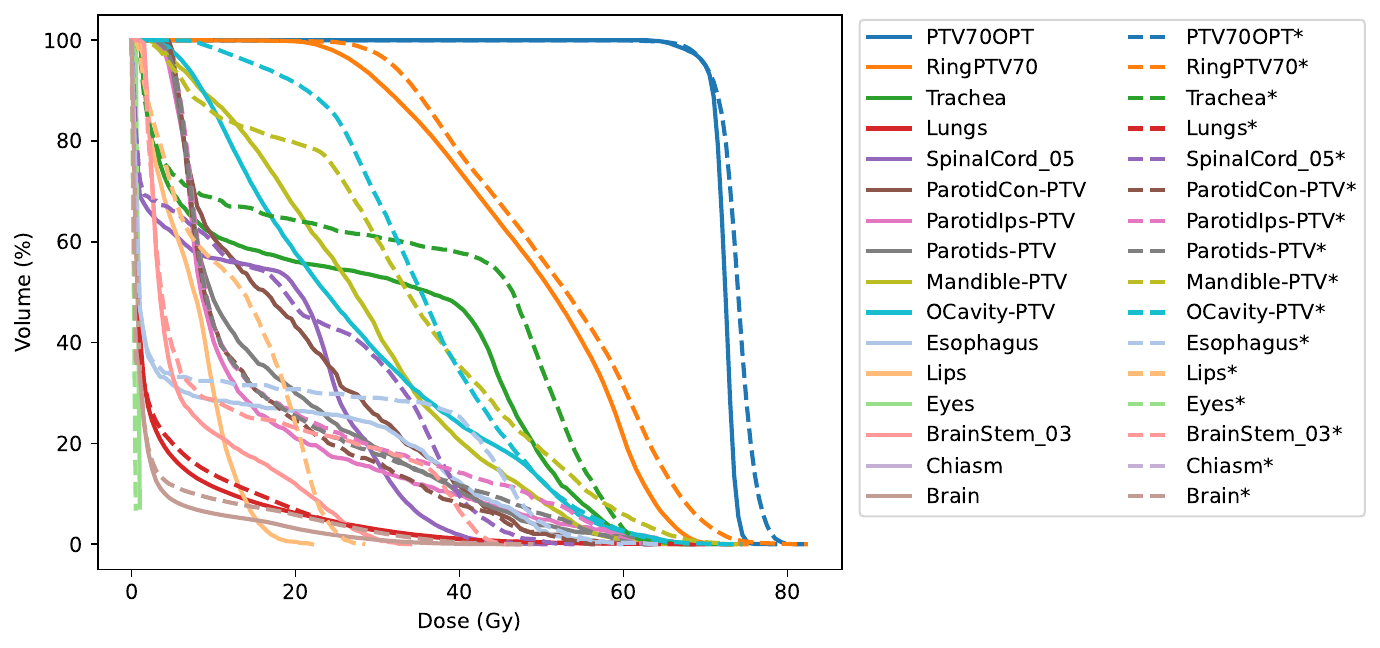}
    \label{fig:second}
\end{subfigure}
\vspace{-0.1in}
\caption{The DVH comparisons between our plans and clinical plans are shown, with dashed lines representing clinical plans and solid lines representing plans generated by our AIRTP pipeline.}
\label{fig:vsclinic}
\end{figure}

\textbf{Comparison with Clinical Plans}. Two typical head-and-neck cases from different cohorts are illustrated in Fig. \ref{fig:vsclinic}. One notable observation is that the AIRTP plan demonstrates improved PTV homogeneity, without compromising OAR sparing under certain scorecards. However, it is important to emphasize that we cannot definitively claim that the AIRTP plan is superior to the clinical manual plan based solely on DVHs or scorecards, as the planning process is inherently subjective and clinical plans may follow different protocols. On a positive note, we have observed the plan quality from our pipeline is high (see more examples in Appendix \ref{sec:supp_moreexp}) following a scorecard definition. This fully automated AIRTP process eliminates human bias, paving the way for unbiased large-scale AI training.

\subsection{Plan Reproducibility Given 3D Dose}
\label{sec:reproduce}
\begin{figure}
\centering
\begin{subfigure}{0.53\textwidth}
    \includegraphics[width=\textwidth]{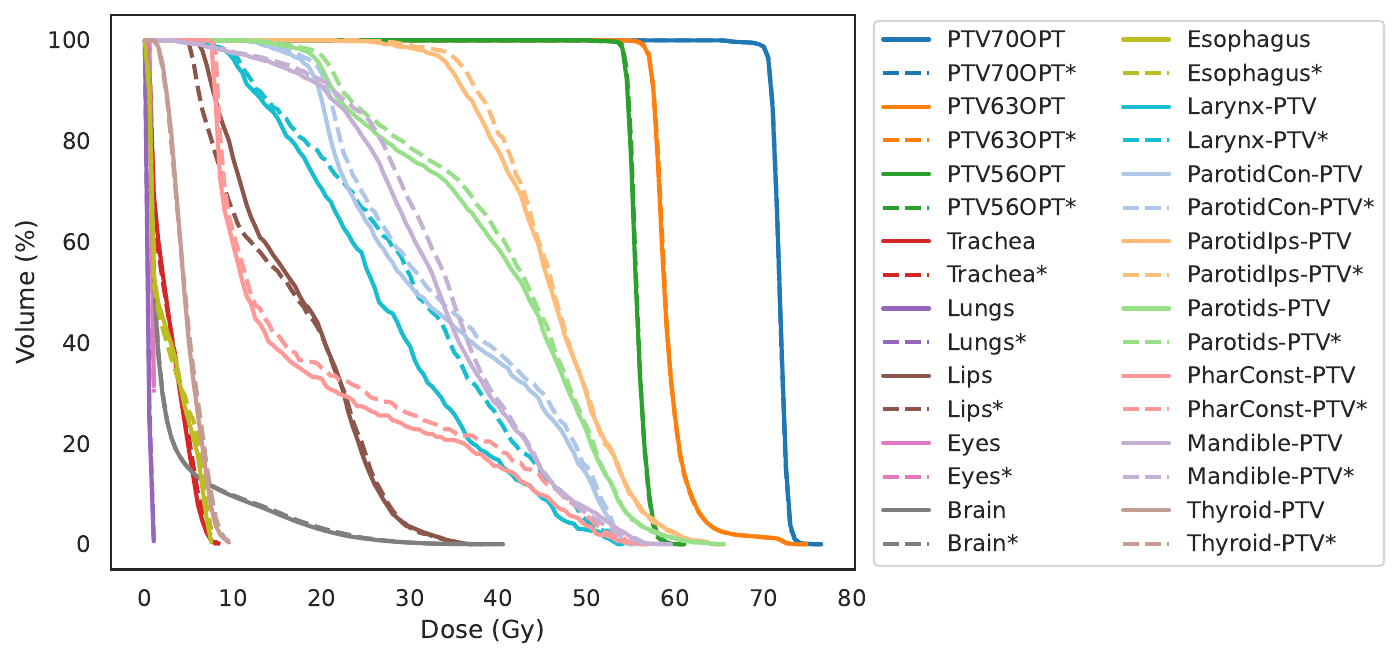}
    \caption{HaN: reference (solid) vs. reproduce (dash). }
    \label{fig:first}
\end{subfigure}
\hfill
\hspace{-1em}
\begin{subfigure}{0.43\textwidth}
    \hspace{-1em}\includegraphics[width=\textwidth]{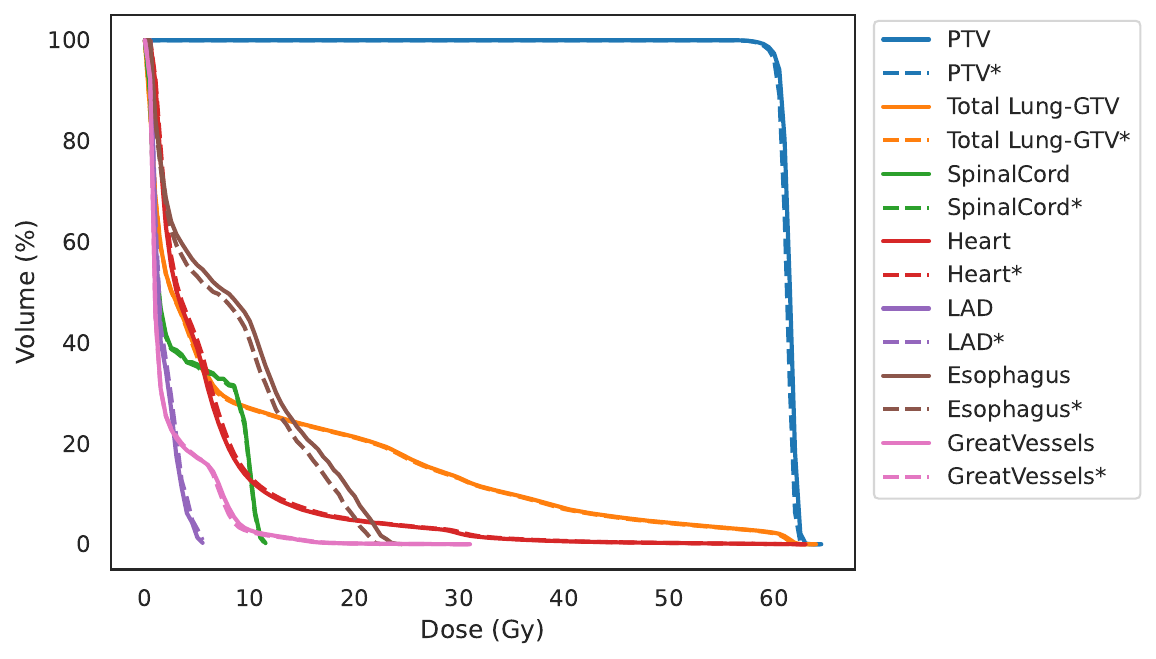}
    \caption{Lung: reference (solid) vs. reproduce (dash).}
    \label{fig:second}
\end{subfigure}
   
\caption{Re-planning examples following the method shown in Fig. \ref{fig:iterativedose}(b) are provided. The closer the dashed line and solid line are to each other, the better the reproducibility of our method.}
\label{fig:replan}
\end{figure}

In Fig. \ref{fig:replan}, we illustrate both head-and-neck and lung cases regarding re-planning consistency based on 3D dose (using the method shown in Fig. \ref{fig:iterativedose}(b)). The PTV objectives are defined according to their prescribed doses to preserve clinical intent, while the OAR objectives are extracted based on the 3D dose of the reference that needs to be reproduced, along with RT contouring.

The advantages of plan reproducibility are twofold. First, when a dose prediction model is properly trained, it enables the rapid generation of visualized 3D dose distributions and DVHs based on patient data and beam configuration. This allows users to quickly assess and adjust the treatment plan. For example, a typical head-and-neck case requires approximately 20 minutes for a single forward-planning session, whereas deep learning-based dose prediction methods, such as those described in \cite{gao2023flexible}, can generate results in just 0.2 seconds (excluding data loading and preprocessing time). Second, this high reproducibility provides an opportunity to evaluate dose predictions in the context of downstream clinical tasks.

\begin{table}[!htb]
    \begin{minipage}{.45\linewidth}
    \renewcommand{\arraystretch}{1.1}
       \caption{Data splits (train/valid/test).}
       \small
        \centering
      \begin{tabular}{ccc}
    \toprule
     &\# Plans & \# Patients  \\
     \midrule
       \cellcolor{Gray}{HaN}  & \cellcolor{Gray}{1428 / 224 / 298}  & \cellcolor{Gray}{499 / 77 / 100}   \\
       Lung  & 1450 / 132 / 200 & 776 / 70 / 100 \\
       \cellcolor{Gray}{Total} & \cellcolor{Gray}{3730} & \cellcolor{Gray}{1622} \\
    \end{tabular}
\label{tab:bs_data}
\end{minipage}%
    \begin{minipage}{.55\linewidth}
    \renewcommand{\arraystretch}{1.1}
     \caption{Results of different phases.}
     \small
        \centering
      \begin{tabular}{lcc}
    \toprule
     & MAE ($\downarrow$) & Plan Quality Score ($\uparrow$) \\
     \midrule
       \cellcolor{Gray}{I: Sanity Check}  & \cellcolor{Gray}1.981  & \cellcolor{Gray}103.09   \\
       II: Validation  & 2.540 & 119.94 \\
       \cellcolor{Gray}III: Testing & \cellcolor{Gray}{2.525} & \cellcolor{Gray}{127.03} \\
    \end{tabular}
\label{tab:bs_res}
    \end{minipage} 
\end{table}

\begin{table}
\centering
\caption{GDP-HMM Challenge test leaderboard sorted by MAE rank}
\begin{tabular}{lccc}
\toprule
\textbf{Team} & \textbf{Solution ID} & \textbf{MAE} & \textbf{MAE Rank} \\
\midrule
Yasin & ID2695 & 2.071 & 1 \\
\rowcolor{gray!10}
rcgao & ID2682 & 2.075 & 2 \\
PVmed & ID2670 & 2.169 & 3 \\
\rowcolor{gray!10}
tyxiong123 & ID2696 & 2.196 & 4 \\
MedVision & ID2673 & 2.208 & 5 \\
\rowcolor{gray!10}
SKLSDE-BH & ID2672 & 2.256 & 6 \\
\bottomrule
\end{tabular}
\label{tab:ch_res}
\end{table}

\subsection{AI benchmarking for Generalizable Dose Prediction}

Many backbones used in dose prediction have adopted architectures originally designed for medical image segmentation \cite{gao2023flexible}.  MedNeXt \cite{roy2023mednext} consistently achieves top performance third-party medical image segmentation TouchStone \cite{bassi2025touchstone} and nnUNet revisit \cite{isensee2024nnu}. Accordingly, we adopt MedNeXt as an educational baseline for generalizable dose prediction using publicly available data from our AIRTP, as illustrated in appended Fig. \ref{fig:ai_backbone}.  The baseline is trained on NVIDIA H100 GPUs. 

Inspired by \cite{gao2023flexible}, we introduce the angle plate and beam plate to encode spatial information related to beam angles. Additionally, categorical information (e.g., treatment mode, anatomical site) is incorporated as an extra input channel. The training dataset is divided into \texttt{dev-train} and \texttt{dev-valid}, where model weights are updated using \texttt{dev-train}, and the best checkpoint is selected based on \texttt{dev-valid}.

To facilitate participation, we provide a complete implementation—including code, data, tutorials, and a pre-trained model—on \href{https://github.com/RiqiangGao/GDP-HMM_AAPMChallenge}{GitHub}, enabling users with minimal radiotherapy background to easily engage with the challenge. An overview of the data distribution is presented in Table \ref{tab:bs_data}, while the baseline results are summarized in Table \ref{tab:bs_res}\footnote{Plan quality scores for validation / testing phases are based on a randomly selected subset of 41 / 50 plans.}. See more about GDP-HMM challenge including detail subcategories data distribution, baseline method, and qualitative results in Appendix  \ref{app:challenge}. 

Table \ref{tab:ch_res} shows the top solutions during the challenge competition. Their source code, trained models, and short technical report can be found in \href{https://huggingface.co/Jungle15/GDP-HMM_baseline}{HuggingFace repository}.  

\subsection{Beyond the Topics of Generalizable Dose Prediction}
\label{sec:othertopic}
\begin{figure}
    \centering
    \includegraphics[width=0.9\linewidth]{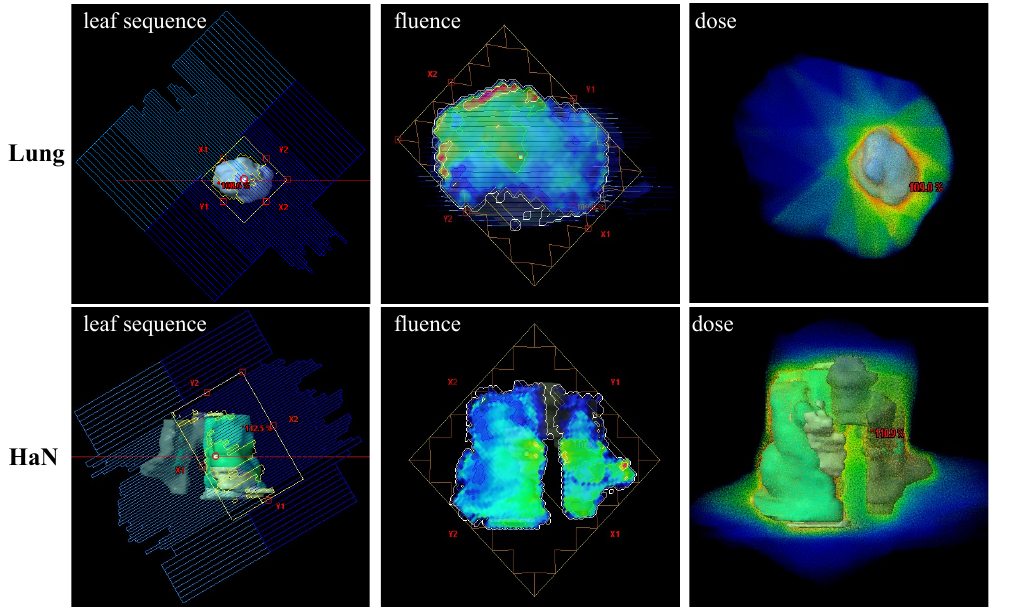}
    \caption{Examples (Visualization from Varian's Eclipse) of DICOM data supporting AI training for RT tasks such as leaf sequencing, fluence prediction, dose calculation, and/or their combinations.}
    \vspace{-0.2in}
    \label{fig:otherapp}
\end{figure}

Beyond dose prediction, researchers have explored AI applications such as fluence prediction \cite{Wang2020FluenceTherapy, arberet2025beam}, leaf sequencing \cite{gao2024multi, hrinivich2024clinical}, and dose calculation \cite{xing2020feasibility,xiao2022transdose}. Integrating one or more of these components can enable end-to-end radiotherapy (RT) planning, potentially advancing toward the next generation of fast and precise AI-driven RT. Our pipeline facilitates large-scale generation of training data across all these modules. As shown in Fig. \ref{fig:otherapp}, we present examples of fluence maps, leaf sequences, and dose volumes generated using our pipeline. The released \href{https://huggingface.co/datasets/Jungle15/Radiotherapy_HaN_Lung_AIRTP}{DICOM datasets} are publicly available to support research in these areas, and a \href{https://github.com/RiqiangGao/GDP-HMM_AAPMChallenge/blob/main/DICOM2NPZ.py}{Python toolkit} is provided to assist with data processing.

\section{Discussion} 
\label{sec:discussion}
\textbf{Conclusion}. In this work, we present an automated radiotherapy planning pipeline that integrates OAR auto-contouring, scripted RT configuration and objective setup, scripted optimization and dose calculation, and automated iterative refinement guided by quality evaluation. The first data release from our AIRTP pipeline offers \textbf{more than 10 times} the number of plans compared to the largest existing well-curated public dataset. Our released dataset can support a wide range of research topics within the RT domain. To promote broader research, we are hosting an AAPM challenge focused on dose prediction task with processed data, source code, tutorials, and pretrained model. We believe that AIRTP, along with the released datasets and associated research activities, has the potential to significantly advance the application of artificial intelligence in radiotherapy. 

\textbf{Limitation and Future Work}. As an initial step toward automating high-quality planning at scale, our work has several limitations and planned future works. \textit{\textbf{First}}, clinical planning involves detailed patient-specific considerations, such as variations in the contouring of PTVs/OARs, auxiliary structures, and various dose constraints. In our pipeline, we focus primarily on clinical PTVs and prescribed doses as the key patient-specific factors, simplifying the process due to the lack of access to more detailed patient data. \textit{\textbf{Second}}, different institutions often follow different protocols, and even within the same institution, planners may have individual preferences. Our pipeline does not currently account for such subjective variations. We believe these preferences can be reflected in the scorecards, here we use only one standardized scorecard per treatment site to scale up for AI model training and evaluation. A future direction is to collaborate with dosimetrists to develop multiple scorecards that better capture diverse clinical preferences and practices. \textit{\textbf{Third}}, unlike fields that benefit from abundant public data and open sourcing, our comparative analyses of iterative refinement and re-planning strategies are limited by the lack of relevant public benchmarking. Although we made efforts to compare the DVHs of AIRTP plans with clinically approved plans, this comparison may not be entirely fair, as the AIRTP and clinical plans follow different guidelines. \textit{\textbf{Fourth}}, this paper focuses on 1) novel methodology to automate high-quality planning at scale, 2) analyses of planning results, 3) open sourcing high-quality data/tutorial/code/pre-trained model for an AAPM challenge and other RT research topics as shown in Sec. \ref{sec:othertopic}. Here we target at AI modeling rather than clinical impact analysis. 

\textbf{Outlook}. Compared to other AI domains such as natural language processing and computer vision, the availability of large-scale training data in medical imaging is often limited. In radiotherapy, this challenge is further amplified by the complexity, subjectivity, noise, and limited scale of clinical data, as discussed in Sec. \ref{sec:rtchallenge}. Despite these obstacles, our work represents a significant step toward enabling large-scale AI research in radiotherapy. Notably, our pipeline is designed to be flexible, making it possible to simulate data at scale with sufficient computational resources and time. We envision this pipeline as a foundational tool for advancing AI applications in radiotherapy and facilitating future research in this critical field with more required data. 

\textbf{Disclaimer}. The information in this paper is based on research results that are not commercially available. Future commercial availability cannot be guaranteed. The released data are provided as-is. We may continuously update tools in the pipeline internally, but the updated data may not be released.

\bibliographystyle{abbrv}  
\bibliography{references}  
\newpage

\appendix

\addcontentsline{toc}{section}{Appendix} %
\renewcommand \thepart{} %
\renewcommand \partname{}
\part{\Large{\centerline{Appendix}}}
\parttoc 

\newpage

\newpage
\section{Data Scale of Popular Successfully AI Models and RT Planning Studies}
\label{sec:data_scale}

We provide additional descriptions for the models presented in Fig. \ref{fig:data_scale}:

\begin{itemize}
    \item \textbf{Stable Diffusion} \cite{rombach2022high}: A state-of-the-art latent diffusion model for generating high-quality images from text prompts, widely adopted for AI-driven image synthesis.

\item \textbf{Segment Anything Model (SAM)} \cite{kirillov2023segment}: A foundational vision model capable of segmenting objects in images with minimal user input, facilitating versatile and efficient image annotation.

\item \textbf{Medical SAM (Med-SAM)} \cite{ma2024segment}: A specialized version of SAM fine-tuned for medical imaging tasks, achieving high-accuracy segmentation across diverse modalities.

\item \textbf{RETFound} \cite{zhou2023foundation}: A foundation model for retinal imaging that employs self-supervised learning to improve disease detection and diagnosis in ophthalmology.

\item \textbf{Medical AI for Synthetic Imaging (MAISI)} \cite{guo2024maisi}: A generative AI framework designed to synthesize realistic medical images for training and validation, enhancing model generalization and robustness.

\item \textbf{Dose Prediction Models} \cite{gao2023flexible, Babier2020OpenKBP:Challenge, fu2024md, kearney2018dosenet}: AI models developed to predict radiation therapy dose distributions, as same as scope in our GDP-HMM challenge.

\item \textbf{Fluence Prediction Models} \cite{Wang2020FluenceTherapy, lee2019fluence}: Computational models that predict fluence maps in radiation therapy, aiding in the optimization of beam intensity for precise dose delivery.

\item \textbf{Leaf Sequencing Models} \cite{gao2024multi, hrinivich2024clinical}: Algorithms that convert optimized fluence maps into deliverable multi-leaf collimator (MLC) sequences, enabling efficient and accurate radiation therapy treatment.

\end{itemize}

Many successful AI models have been trained on large-scale, carefully curated datasets. In contrast, applying AI in radiation therapy (RT) planning faces several major challenges: (1) available training samples are often very limited, typically fewer than 1,000 cases; (2) clinical RT data frequently contain missing or noisy information; and (3) most high-quality curated RT datasets are not publicly available.
\textbf{\textit{Motivated by the goal of advancing RT planning with AI, this study has been conducted to address these critical challenges.}}

\newpage
\section{Challenges of Clinical Plans for AI Training}

\begin{table}[h]
    \scriptsize 
    \centering 
    \caption{The heterogeneity of RT structure naming. There are thousands of unique names for PTVs and OARs across cohorts due to varied convention, making data curation difficult for AI training.}
    \begin{tabular}{ccccl}
    \toprule
    Cohort & \# Patients & ROI &\# Unique Names & Name Examples \\ 
\midrule
\cellcolor{Gray}{In-house cohort} & \cellcolor{Gray}702 & \cellcolor{Gray}PTV & \cellcolor{Gray}336 & 
\cellcolor{Gray}\parbox{7cm}{\texttt{PTV5000, ptv inside, 3PTV\_5600, 2PTV w ON, PTVplanning, PTV 70, 2PTV\_4800\_Uniform, 2PTV\_P, ptv-temp, 1ptv w, 1ptv inf, PTV46 Original, 1left PTV }} \vspace{0.05in}\\ 
 & & OAR & 2597 & 
\parbox{7cm}{\texttt{Brain1, Brain\_partial, BrainStem\_PRV, Cord High + .5, Cord, LarynxGSL, Left BP, OC, OC uninvolved, OC wo, OC-lips, OC\_Lips, ON\_03, OralCavity, OralCavity\_JJC, OralCavity with}}  \vspace{0.05in}\\ 
\cellcolor{Gray}{TCIA (5 cohorts)} & \cellcolor{Gray}690 & \cellcolor{Gray}PTV & \cellcolor{Gray}1539 & 
\cellcolor{Gray}\parbox{7cm}{\texttt{5445 PTV, COLD PTV6996, FINAL PTV69, MIN PTV66, Opti\_PTV60, PTV 60 OPTI\_AP\_0\_AP, PTV 60 old, PTV 66 Sup, PTV EXP, PTV HD, PTV NECK, PTV PAROTIDES}}  \vspace{0.05in}\\ 
 & & OAR & 3694 & 
\parbox{7cm}{\texttt{Brain DNU, Brain Push, Brachial Plexus, Brain sub, Brain2, Brainstem expd, Bstem\_PRV, Cord + 0.5cm, Cord$<$ 45, Cord\_PRV, Cord+3, Cord+5mm}}\\
\end{tabular}
    \label{tab:rtnames}
\end{table}

\begin{figure}[h]
    \centering
    \includegraphics[width=0.8\linewidth]{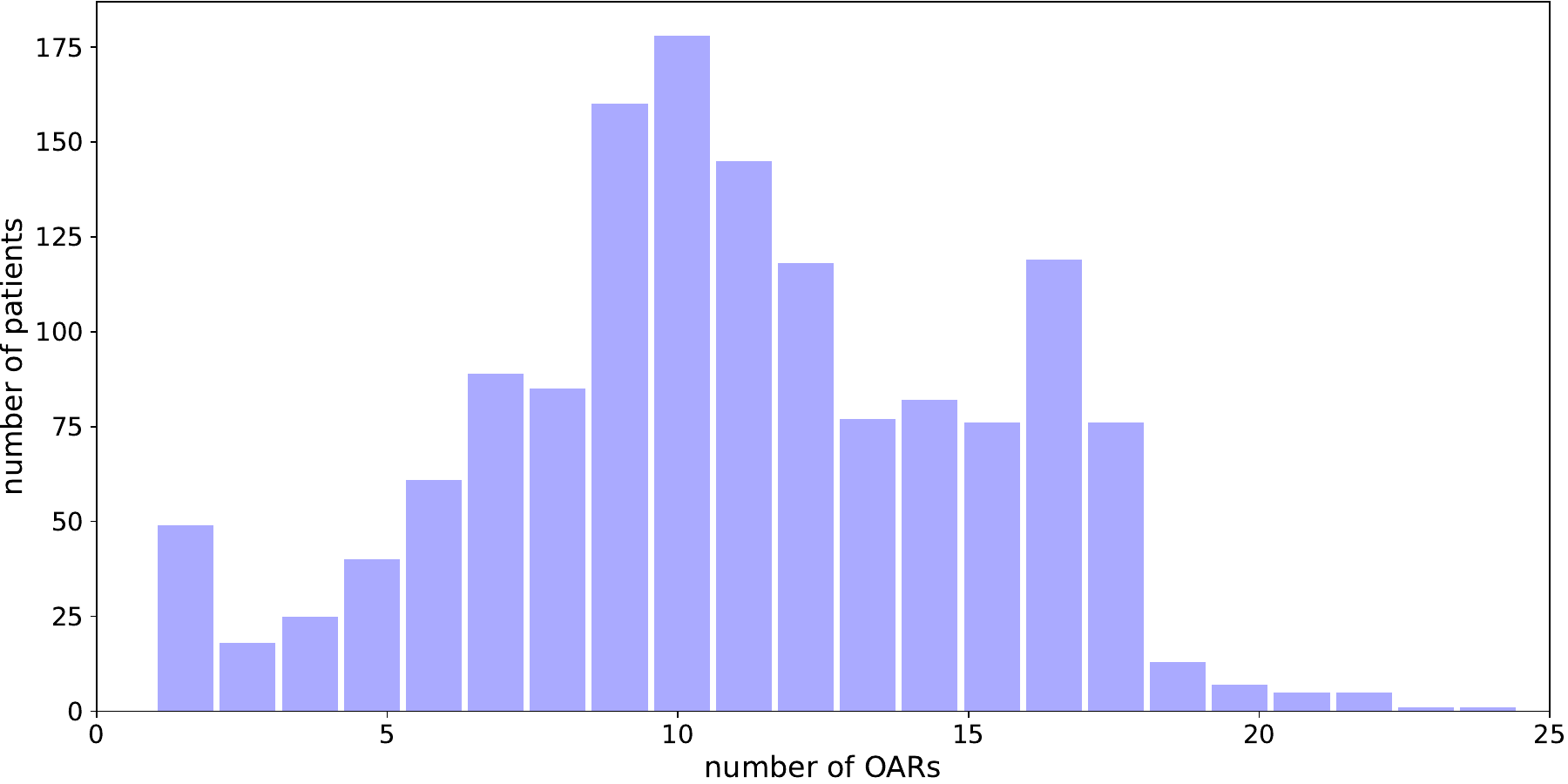}
	\caption{\small Number of clinically contoured OARs varies significantly across patients.}
\label{fig:oar_cnt}
\end{figure}

As shown in Table \ref{tab:rtnames}, the naming of RT structures is highly heterogeneous. Naming conventions often vary across institutions, annotators, and even individual patients. Curating such datasets for AI training typically requires extensive manual review, despite the possibility of implementing some rule-based approaches. The presence of thousands of different — and sometimes subjective — structure names makes curation inefficient and severely limits the scalability of training data.

Fig. \ref{fig:oar_cnt} further illustrates that the number of clinically contoured OARs varies significantly across patients, adding another layer of complexity for AI model development. In addition, the optimization objectives used to create clinical treatment plans are usually not recorded. These objectives are often complex, and their association with RT structures is not transparent, as some RT structures may not be included in the objectives.

While limited-scale datasets have enabled AI success in relatively simple scenarios — as referenced in Appendix \ref{sec:data_scale} — they may require considerable effort in data curation. Building generalizable and robust AI models for real-world clinical settings remains a significant challenge. We believe that our work on preparing diverse clinical contexts and generating high-quality treatment plans at scale will be crucial for advancing the next generation of AI in radiotherapy planning.

\newpage
\section{Standardized Structures and Score Card Items}
\label{app:structures}
\begin{table}[h]
\scriptsize
    \centering
     \renewcommand{\arraystretch}{1.3}
     \caption{HaN RT structures for RapidPlan \cite{magliari2024hn} and scorecard \cite{varian2024han}.}
    \begin{tabular}{llll}
    \toprule
     \parbox{3cm}{\textbf{PTV \& Helpers}}  & \parbox{3cm}{Comments} & \parbox{3cm}{\textbf{PTV \& Helpers}}  & \parbox{3cm}{Comments} \\
     \midrule
      \cellcolor{Gray}\texttt{PTVHigh}  & \cellcolor{Gray}\parbox{3cm}{\fontsize{6}{6}\selectfont{PTV with the largest prescription}}& \texttt{RingPTVHigh} & \parbox{3cm}{\fontsize{6}{6}\selectfont{[(\texttt{PTVHigh} + 30mm) – (\texttt{PTVHigh} + 2mm)] AND \texttt{Body}}}
  \\
      
      \texttt{PTVHighOPT}  & \parbox{3cm}{\fontsize{6}{6}\selectfont{\texttt{PTVHigh} – (\texttt{BrachialPlexus} + 2mm)}}& \cellcolor{Gray}& \cellcolor{Gray}\\
      
      \cellcolor{Gray}\texttt{PTVMid}  & \cellcolor{Gray}\parbox{3cm}{\fontsize{6}{6}\selectfont{PTV with the middle prescriptions}}& \texttt{RingPTVMid} & \parbox{3cm}{\fontsize{6}{6}\selectfont{ [(\texttt{PTVMid} + 30mm) – (\texttt{PTVMid} + 2mm) - (\texttt{PTVHigh} + 6mm)] AND \texttt{Body} }}\\
      
      \texttt{PTVMidOPT}   & \parbox{3cm}{\fontsize{6}{6}\selectfont{\texttt{PTVMid} – (\texttt{PTVHigh} + 3mm)}}& \cellcolor{Gray}\texttt{PTVMid-PTVHigh} &  \cellcolor{Gray}\fontsize{6}{6}\selectfont{as the structure name} \\

      \cellcolor{Gray}\texttt{PTVLow}  & \cellcolor{Gray}\parbox{3cm}{\fontsize{6}{6}\selectfont{PTV with the lowest prescription}}& \texttt{RingPTVLow} & \parbox{3cm}{\fontsize{6}{6}\selectfont{[(\texttt{PTVLow} + 30mm) – (\texttt{PTVLow} + 2mm) - (\texttt{PTVMid} + 6mm) - (\texttt{PTVHigh} + 9mm)] AND \texttt{Body}}}  \\
      
      \texttt{PTVLowOPT}  & \parbox{3cm}{\fontsize{6}{6}\selectfont{\texttt{PTVLow} - (\texttt{PTVMid} + 3mm ) - (\texttt{PTVHigh} + 6mm)}}& \cellcolor{Gray} \texttt{PTVLow-PTVMid} & \cellcolor{Gray}\fontsize{6}{6}\selectfont{as the structure name}  \\

    \cellcolor{Gray}\texttt{PTVTotal} & \cellcolor{Gray}\parbox{3cm}{\fontsize{6}{6}\selectfont{union of all PTVs}}& &  \\
    
    \toprule
     \textbf{OARs} & \textbf{OARs} & \textbf{OARs} & \textbf{OARs-PTV} \\
     \midrule
     \cellcolor{Gray}\texttt{BrachiaPlexus} & \texttt{Larynx} & \cellcolor{Gray}\texttt{Posterior\_Neck}& \texttt{Larynx-PTV} \\
     
     \texttt{Brain} & \cellcolor{Gray}\texttt{Lens\_L} & \texttt{Shoulders}& \cellcolor{Gray} \texttt{Mandible-PTV} \\
     
     \cellcolor{Gray}\texttt{BrainStem} & \texttt{Lens\_R} & \cellcolor{Gray}\texttt{SpinalCord}& \texttt{OCavity-PTV} \\
     
     \texttt{BrainStem\_03} & \cellcolor{Gray}\texttt{Lips} &\texttt{SpinalCord\_05} & \cellcolor{Gray}\texttt{ParotdCon-PTV} \\
     
     \cellcolor{Gray}\texttt{Chiasm} & \texttt{OpticNerve\_L}  &\cellcolor{Gray} \texttt{Submandibular} & \texttt{ParotdIps-PTV} \\
     
     \texttt{Cochlea\_L} & \cellcolor{Gray}\texttt{OpticNerve\_R}& \texttt{Trachea}& \cellcolor{Gray}\texttt{Parotids-PTV} \\
     
     \cellcolor{Gray}\texttt{Cochlea\_R} & \texttt{OralCavity}& \cellcolor{Gray}\texttt{Lungs}& \texttt{PharCost-PTV} \\
     
     \texttt{Esophagus} & \cellcolor{Gray}\texttt{Parotids}& \texttt{Mandible} & \cellcolor{Gray}\texttt{Submand-PTV} \\
     \cellcolor{Gray}\texttt{Eyes} & \texttt{PharynxConst} & \cellcolor{Gray}\texttt{Body} & \texttt{SubmandL-PTV} \\
     \texttt{LacrimalGlands} & \cellcolor{Gray}\texttt{Pituitary} & \texttt{TMJoint} & \cellcolor{Gray}\texttt{SubmandR-PTV} \\
    \end{tabular}
    \label{tab:contours}
\end{table}

Table \ref{tab:contours} presents original RapidPlan \cite{magliari2024hn} contouring items in head-and-neck cancer. Some organ contours may be missing due to several reasons: (1) the regions extend beyond the CT field of view, (2) there is significant overlap with clinical PTVs, or (3) segmentation failures occur, such as zero or extremely small segmented masks, (4) coustomized scorecard may not include all the structures.  Additionally, as noted in the limitations, variations in contouring guidelines and individual planner preferences can introduce subjectivity. Simply combining clinical PTVs with auto-generated OAR masks and helper structures may not directly translate to clinical practice, as manual adjustments by planners are often necessary. While we cannot guarantee that every large-scale planning solution meets clinical standards, this approach still provides substantial value for AI model development, as noted in the main text. 

\newpage
\begin{table}[h]
    \scriptsize
    \begin{minipage}{.6\linewidth}
    \renewcommand{\arraystretch}{1.3}
      \raggedright
       \caption{Initial lung optimization objectives.}
      \begin{tabular}{llllll}
     \toprule
     ROI & Type & Dose & Volume & Operation & Prior\\  
     \midrule
    \cellcolor{Gray}\texttt{PTV} & \cellcolor{Gray}Point & \cellcolor{Gray}60 Gy & \cellcolor{Gray}100 \% & \cellcolor{Gray}Lower & \cellcolor{Gray}300   \\
    \texttt{PTV} & Point & 63 Gy & 0 \% & Upper & 300  \\
    \cellcolor{Gray}\texttt{BodyRing} & \cellcolor{Gray}Point & \cellcolor{Gray}63 Gy & \cellcolor{Gray}0 \% & \cellcolor{Gray}Upper & \cellcolor{Gray}250  \\
    \texttt{TotalLung-GTV} & Point & 20 Gy & 18.5 \% & Upper & 100  \\
    \cellcolor{Gray}\texttt{TotalLung-GTV} & \cellcolor{Gray}Point & \cellcolor{Gray}5 Gy & \cellcolor{Gray}32.5 \% & \cellcolor{Gray}Upper & \cellcolor{Gray}100  \\
    \texttt{TotalLung-GTV} & Mean & 5 Gy & N/A & Upper & 100 \\
    \cellcolor{Gray}\texttt{SpinalCord} & \cellcolor{Gray}Point & \cellcolor{Gray}22 Gy & \cellcolor{Gray}0 \% & \cellcolor{Gray}Upper & \cellcolor{Gray}80   \\
    \texttt{Heart} & Point & 55 Gy & 0 \% & Upper & 60   \\
    \cellcolor{Gray}\texttt{Heart} & \cellcolor{Gray}Mean & \cellcolor{Gray}5 Gy & \cellcolor{Gray}N/A & \cellcolor{Gray}Upper & \cellcolor{Gray}60 \\
    \texttt{LAD} & Point & 55 Gy & 0 \% & Upper & 60   \\
    \cellcolor{Gray}\texttt{Esophagus} & \cellcolor{Gray}Mean & \cellcolor{Gray}15 Gy & \cellcolor{Gray}N/A & \cellcolor{Gray}Upper & \cellcolor{Gray}60 \\
    \texttt{BrachPlexus} & Mean & 15 Gy & N/A & Upper & 60 \\
    \cellcolor{Gray}\texttt{GreatVessels} & \cellcolor{Gray}Mean & \cellcolor{Gray}55 Gy & \cellcolor{Gray}0 \% & \cellcolor{Gray}Upper & \cellcolor{Gray}60   \\
    \texttt{Trachea} & Mean & 55 Gy & 0 \% & Upper & 60   \\
    \cellcolor{Gray}\texttt{RingPTV} & \cellcolor{Gray}Mean & \cellcolor{Gray}67 Gy & \cellcolor{Gray}0 \% & \cellcolor{Gray}Upper & \cellcolor{Gray}60   \\
     & & & & \\
\end{tabular}
\label{tab:lungobj}
\end{minipage}%
    \begin{minipage}{.4\linewidth}
    \renewcommand{\arraystretch}{1.17}
     \caption{Lung scorecard items.}
      \begin{tabular}{lll}
     \toprule
     ROI & Metric Type & \parbox{1.2cm}{Input}  \\
     \midrule
    \cellcolor{Gray}\texttt{PTV} & \cellcolor{Gray}VolAtDose & \cellcolor{Gray}60 Gy   \\
    \texttt{CTV} & VolAtDose & 60 Gy   \\
    \cellcolor{Gray}\texttt{PTV} & \cellcolor{Gray}VolAtDose & \cellcolor{Gray}57 Gy \\
    \texttt{PTV} & VolAtDose & 54 Gy   \\
    \cellcolor{Gray}\texttt{PTV} & \cellcolor{Gray}DoseAtVol & \cellcolor{Gray}0.03 cc   \\
    \texttt{PTV} & DoseAtVol & 99.5 \%  \\
    \cellcolor{Gray}\texttt{TotalLung-GTV} & \cellcolor{Gray}VolAtDose & \cellcolor{Gray}20 Gy  \\
    \texttt{TotalLung-GTV} & VolAtDose & 5 Gy  \\
    \cellcolor{Gray}\texttt{TotalLung-GTV} & \cellcolor{Gray}MeanDose & \cellcolor{Gray}N/A  \\
    \texttt{SpinalCord} & DoseAtVol & 0.03 cc   \\
    \cellcolor{Gray}\texttt{Heart} & \cellcolor{Gray}DoseAtVol & \cellcolor{Gray}0.03 cc   \\
    \texttt{Heart} & MeanDose & N/A  \\
    \cellcolor{Gray}\texttt{LAD} & \cellcolor{Gray}VolAtDose & \cellcolor{Gray}15 Gy \\ 
    \texttt{Esophagus} & DoseAtVol & 0.03 cc \\
    \cellcolor{Gray}\texttt{Esophagus} & \cellcolor{Gray}MeanDose & \cellcolor{Gray}N/A \\
    \texttt{BrachPlexus} & DoseAtVol & 0.03 cc \\
    \cellcolor{Gray}\texttt{GreatVessels} & \cellcolor{Gray}DoseAtVol & \cellcolor{Gray}0.03 cc \\
    \texttt{Trachea} & DoseAtVol & 0.03 cc \\
\end{tabular}
\label{tab:lungscard}
    \end{minipage} 
\end{table}

Since a publicly available RapidPlan model that adequately covers the various lung tumor locations is not found, we instead derive the optimization objectives directly from the scorecard for initial lung cancer planning, as shown in Table \ref{tab:lungobj}. The corresponding lung scorecard items are listed in Table \ref{tab:lungscard}.

\newpage

\section{Data Curation Pipeline with AIRTP Integration}
\label{app:curation}
\begin{figure}[h]
    \centering
    \includegraphics[width=0.98\linewidth]{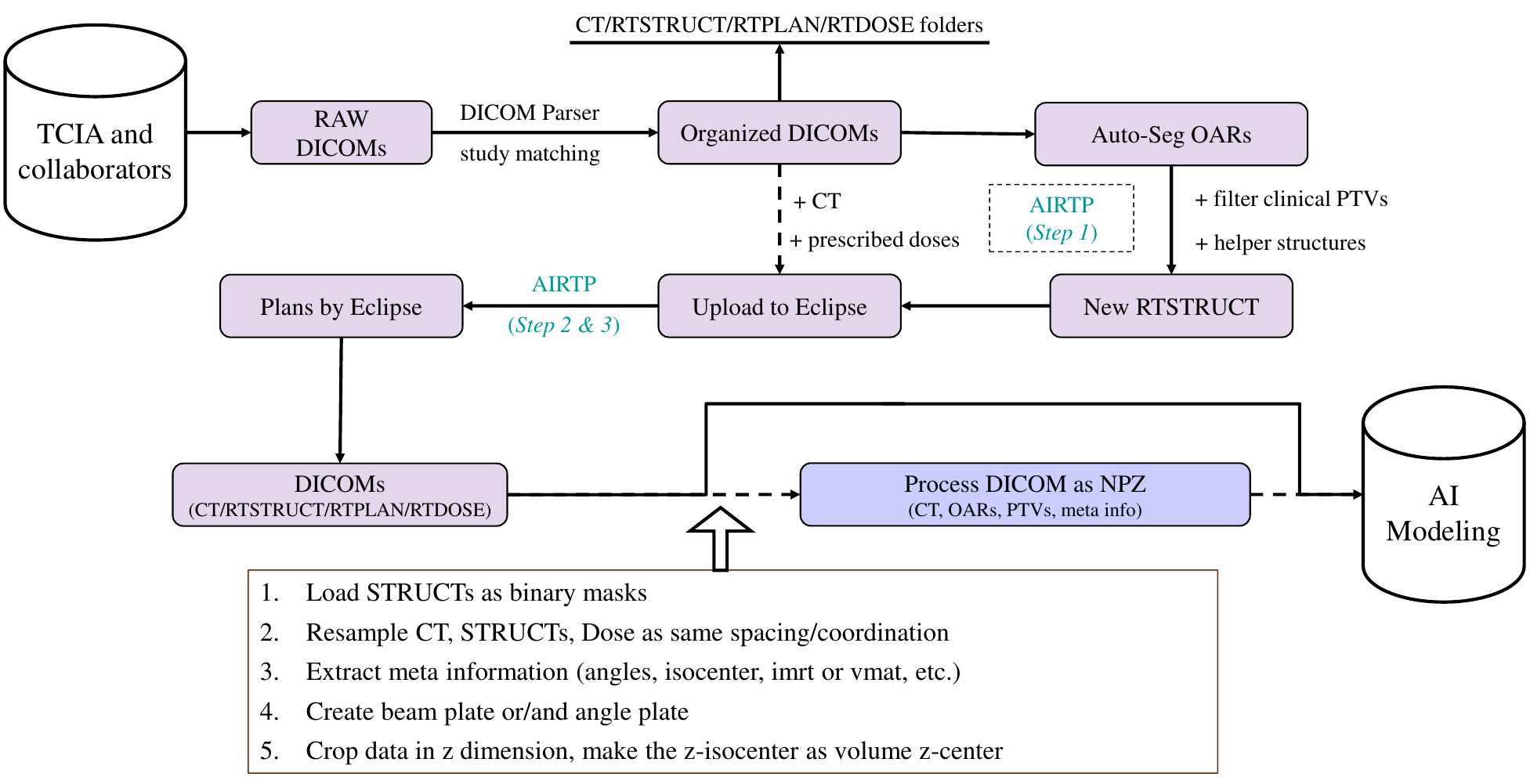}
    \caption{Overview of the data curation pipeline, from raw data processing to AI model training, highlighting where AIRTP integrates into the workflow.}
    \label{fig:curation}
\end{figure}

The complete data curation pipeline is illustrated in Fig. \ref{fig:curation}. Raw data from public archives such as \href{https://www.cancerimagingarchive.net/}{TCIA} or from clinical collaborators are often not well organized. We use a DICOM parser tool to structure each study into separate folders containing CT, RTSTRUCT, RTPLAN (if available), and RTDOSE (if available) data. In Step 1 of the AIRTP pipeline, we segment all OARs, filter clinical PTVs through name and dose-level matching, and create helper structures, following the list provided in Appendix \ref{app:structures}. Along with the CT series, the newly generated RTSTRUCT data are batch-uploaded to the Eclipse server. Using Step 2 and 3 of AIRTP, we then generate high-quality RT plans at scale. Finally, DICOM files for CT, RTSTRUCT, RTPLAN, and RTDOSE are exported locally.

Exported DICOMs may be directly used for AI model training. Alternatively, researchers may process data and save them in more efficient formats (e.g., Numpy) to facilitate data loading. As an example, we provide a \href{https://github.com/RiqiangGao/GDP-HMM_AAPMChallenge/blob/main/DICOM2NPZ.py}{Python toolkit} alongside the hosted GDP-HMM challenge to support data processing.

\newpage

\section{Code Blocks for Algorithm \ref{alg:iter}} 

\begin{figure}[h]
    \centering
    \begin{lstlisting}%[style=mypython]
    
    def Dose2Obj(s, value, volume, score):
        if '-PTV' in s.Id and value > 20:
            res = max(value * 0.95, value - 3)
            weight = 120
        elif value > 63 and '-PTV' not in s.Id:
            res = value - 1
            weight = 80
        elif value < 20 and (volume is not None and volume < 50):
            res = max(value * 0.95, value - 2)
            weight = 80
        elif value < 10:
            res = max(value * 0.95, value - 2)
            weight = 80
        else:
            res = max(value * 0.95, value - 2)
            weight = 120
        return round(res, 2), weight
\end{lstlisting}
    \caption{Calculation of optimization objectives based on DVHs during the iterative process.}
    \label{code:dose2obj}
\end{figure}

\begin{figure}[h]
    \centering
    \begin{lstlisting}%[style=mypython]
    
    def scorecard2dose(x, ScorePoints):
        ScorePoints = sorted(ScorePoints, key=lambda p: p['PointX'])
    
        if x < ScorePoints[0]['PointX']:
            x =  ScorePoints[0]['PointX']
        if x > ScorePoints[len(ScorePoints) - 1]['PointX']:
            x = ScorePoints[len(ScorePoints) - 1]['PointX']
        
        for i in range(len(ScorePoints) - 1):
            x1, y1 = ScorePoints[i]['PointX'], ScorePoints[i]['Score']
            x2, y2 = ScorePoints[i + 1]['PointX'], ScorePoints[i + 1]['Score']
            assert x1 <= x2
            # Check if x is within the current segment
            if x1 <= x <= x2:
                # Calculate the slope (m) and intercept (b) for the line between (x1, y1) and (x2, y2)
                m = (y2 - y1) / (x2 - x1 + 1e-5)
                b = y1 - m * x1
                score = m * x + b
                dose_value = max(x * 0.85, x - 7) # add margin to dose value
                priority = 60 + min(abs(m), 7) * 20 / (1 + np.exp(score / 3)) 
                # calculate the optimization priority
                return dose_value, priority
\end{lstlisting}
    \caption{Calculation of optimization objectives using scorecard items during the iterative process. }
    \label{code:scorecard2dose}
\end{figure}

\newpage
\section{More about GDP-HMM Challenge}
\label{app:challenge}

\subsection{Data Distribution}
\begin{table}[h]
    \centering
    \caption{Data Distribution for GDP-HMM Challenge.}
    \footnotesize
    \label{tab:distri}
    \begin{tabular}{lllll}
    \toprule
        &   & Training Split    & Validation Split   & Test Split    \\ 
\midrule
\cellcolor{Gray}\#Plans  & \cellcolor{Gray} &  \cellcolor{Gray}  &  \cellcolor{Gray} &  \cellcolor{Gray}   \\ 
 & HaN    &  1426 (49.6\%) &  224 (62.9\%) &  298 (59.8\%)  \\ 
 & Lung    &  1450 (50.4\%) & 132 (37.1\%) &  200  (40.2\%)  \\ 
\cellcolor{Gray}\#Patients  & \cellcolor{Gray} &  \cellcolor{Gray}  &  \cellcolor{Gray} &  \cellcolor{Gray}  \\ 
 & HaN    &  498 (39.1\%) & 77 (52.4\%)  &   100 (50.0\%)  \\ 
 & Lung    &  776 (60.9\%) & 70 (47.6\%) &  100  (50.0\%) \\ 
\cellcolor{Gray}\#Plan by \#PTV  & \cellcolor{Gray} &  \cellcolor{Gray}  &  \cellcolor{Gray} &  \cellcolor{Gray}   \\ 
 & 3PTVs per Plan (HaN) &  912 (31.7\%) & 125 (35.1\%)  &   148 (29.7\%)  \\ 
 & 2PTVs per Plan (HaN)  &  413 (14.4\%) & 84 (23.6\%) &  120  (24.1\%) \\ 
 & 1PTV per Plan (HaN)  &  101 (3.5\%) & 15 (4.2\%) &  30  (6.0\%) \\
 & 1PTV per Plan (Lung)  &  1450 (50.4\%) & 132 (37.1\%) &  200  (40.2\%) \\
 \cellcolor{Gray}\#Plans by Type  & \cellcolor{Gray} &  \cellcolor{Gray}  &  \cellcolor{Gray} &  \cellcolor{Gray}  \\ 
 & IMRT    &  1645 (57.2\%) &  212 (59.6\%) &  300 (60.2\%)  \\ 
 & VMAT    &  1231 (42.8\%) & 144 (40.4\%) &   198  (39.8\%)  \\
\end{tabular}
\end{table}

\begin{figure}[h]
\centering

\begin{subfigure}{0.49\textwidth}
    \includegraphics[width=\textwidth]{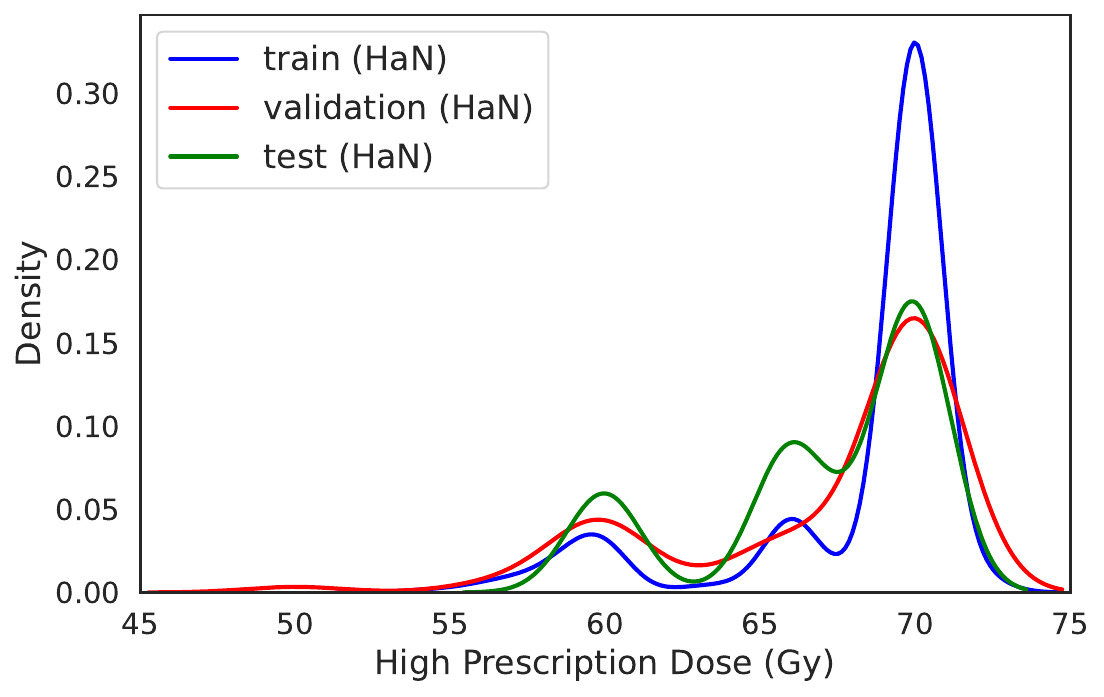}
    \label{fig:first}
\end{subfigure}
\hfill
\begin{subfigure}{0.48\textwidth}
    \includegraphics[width=\textwidth]{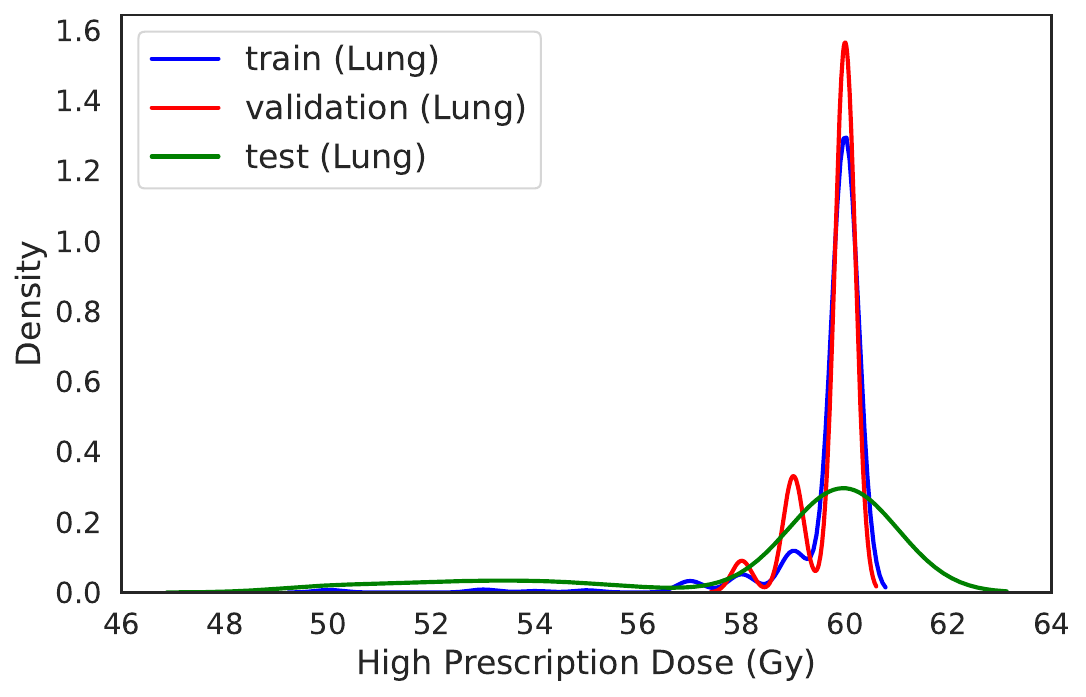}
    \label{fig:second}
\end{subfigure}
\caption{Kernel density estimation of the high prescription dose in different data splits.}
\label{fig:pdose}
\end{figure}

The data distribution of GDP-HMM challenge is shown in Table \ref{tab:distri} and Fig. \ref{fig:pdose}. The testing patients are from different institutes than the training and validation splits.  
\newpage
\subsection{Baseline Method}

\begin{figure}[h]
    \centering
    \includegraphics[width=0.9\linewidth]{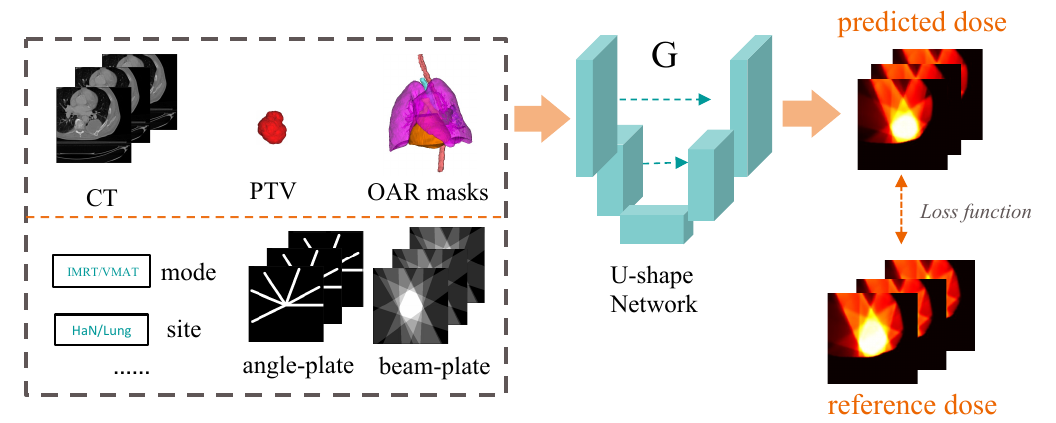}
    \caption{A baseline for the generalizable dose prediction for GDP-HMM challenge. The MedNeXt is used as backbone (U-shape Network).}
    \label{fig:ai_backbone}
\end{figure}

As shown in Fig. \ref{fig:ai_backbone}, the educational baseline applies the MedNeXt as the backbone; due to its superior performance in multiple large-scale segmentation benchmarks. Fundamentally, the dose prediction is an conditional image generation problem. The CT images, PTV \& OAR masks, planning mode, treatment site, and beam geometries etc., are conditions for 3D dose prediction. We follow the strategy of from \cite{gao2023flexible}, concatenating all the conditions as multi-channel 3D input. The loss function used in the baseline is pretty simple: the mean absolute error (MAE) of the predicted dose and the reference dose. 

For more details, please refer to links of \href{https://github.com/RiqiangGao/GDP-HMM_AAPMChallenge}{code \& tutorial} and \href{https://huggingface.co/Jungle15/GDP-HMM_baseline}{pre-trained model}. 

\newpage

\subsection{Result Samples from the Baseline}

\begin{figure}[h]
    \centering
    \includegraphics[width=0.90\linewidth]{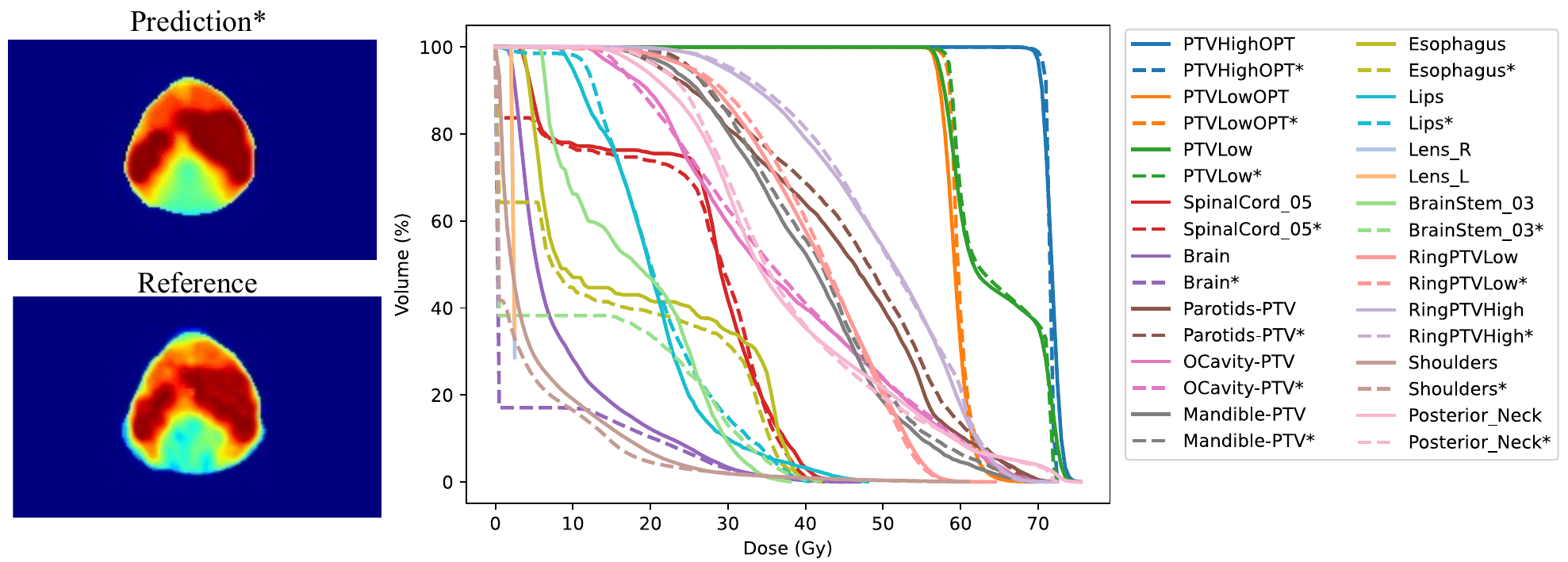}
    \caption{Head-and-neck IMRT example.}
    \label{fig:han_imrt}
\end{figure}
\begin{figure}[h]
    \centering
    \includegraphics[width=0.90\linewidth]{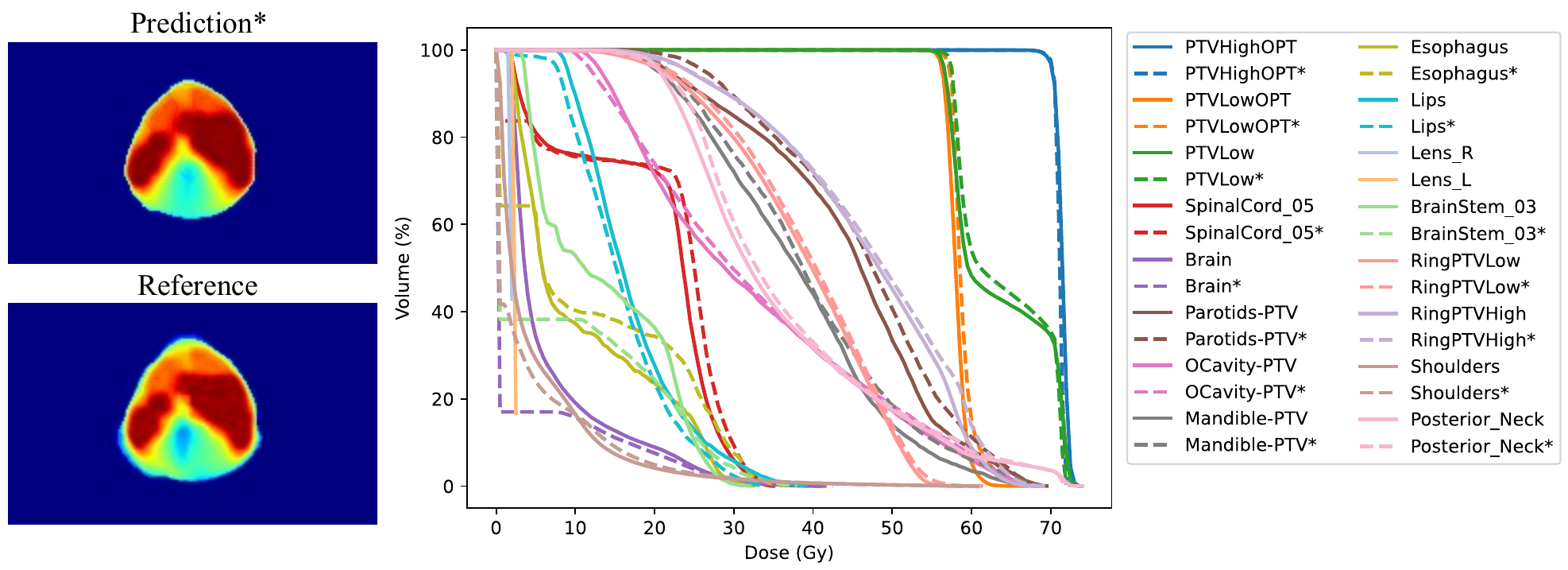}
    \caption{Head-and-neck VMAT example.}
    \label{fig:han_vmat}
\end{figure}

\begin{figure}[h]
    \centering
    \includegraphics[width=0.75\linewidth]{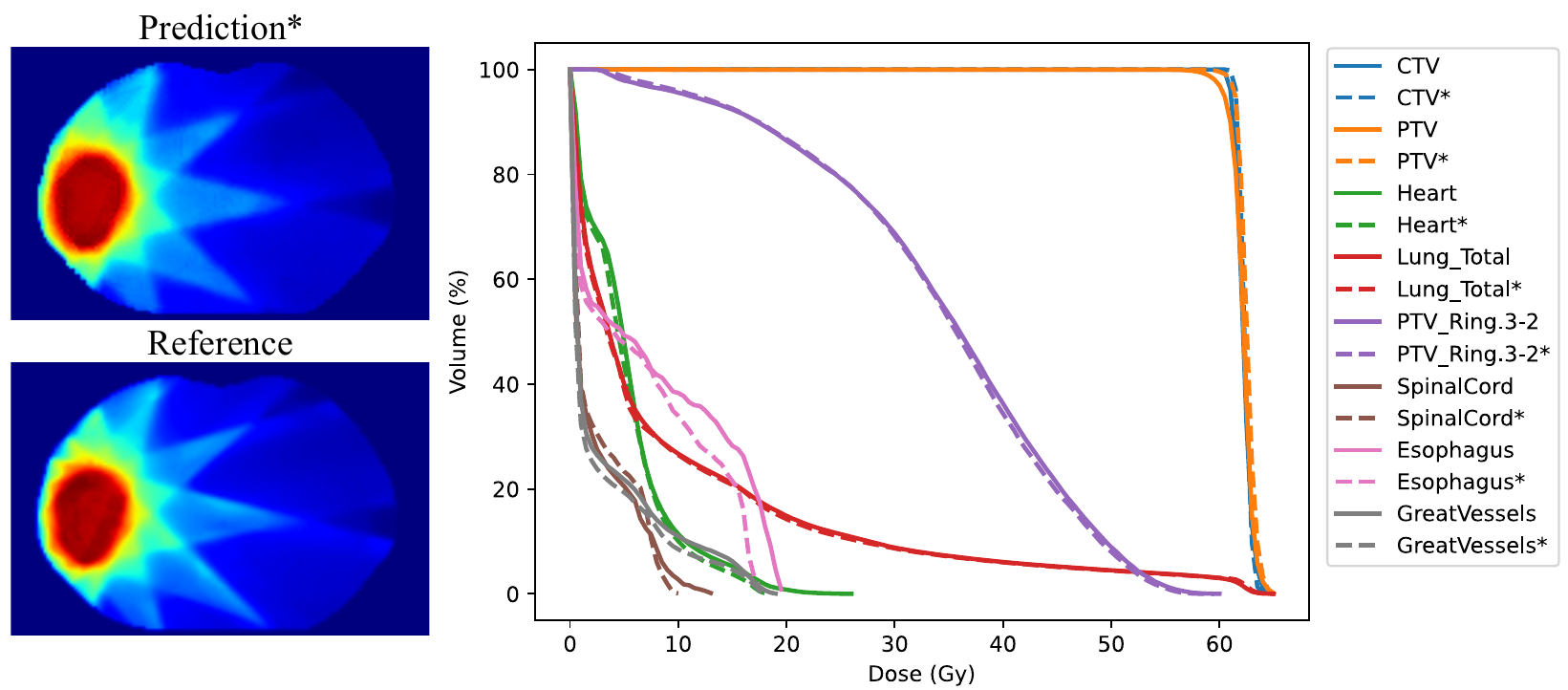}
    \caption{Lung IMRT example.}
    \label{fig:lung_imrt}
\end{figure}

\begin{figure}[h]
    \centering
    \includegraphics[width=0.75\linewidth]{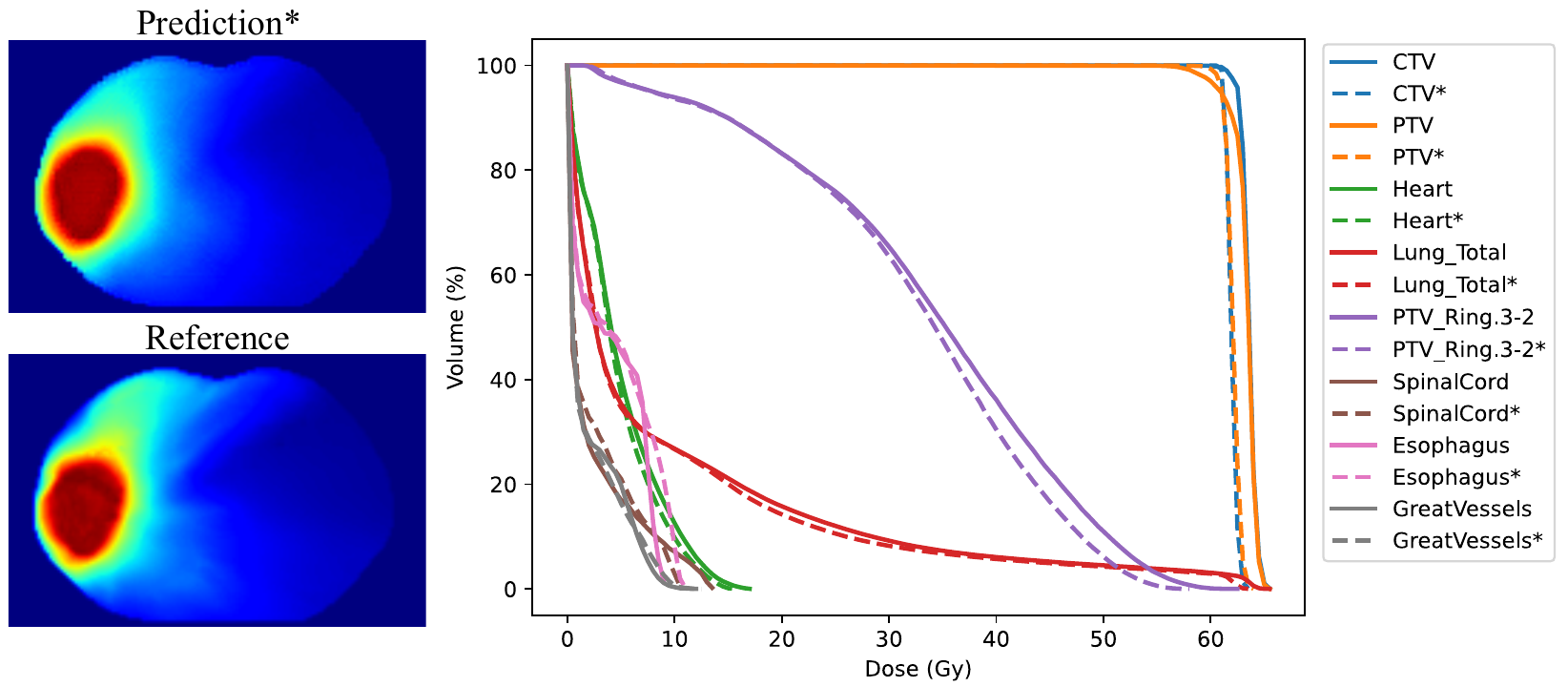}
    \caption{Lung VMAT example.}
    \label{fig:lung_vmat}
\end{figure}

We present four samples from our validation cohort, illustrating distinct scenarios that a universal AI model should be capable of handling. Fig. \ref{fig:han_imrt} and \ref{fig:han_vmat} depict treatment plans for the same head-and-neck cancer patient using different modalities: IMRT and VMAT. Similarly, Fig. \ref{fig:lung_imrt} and \ref{fig:lung_vmat} show treatment plans for the same lung cancer patient, also with IMRT and VMAT. Each case includes a comparison of the predicted and reference 3D dose distributions, along with corresponding dose-volume histograms (DVHs). 

\newpage

\newpage

\section{Visualization of Creating Beam/Angle Plates}
\label{sec:supp_plate}
\begin{figure}[h]
    \centering
    \includegraphics[width=0.9\linewidth]{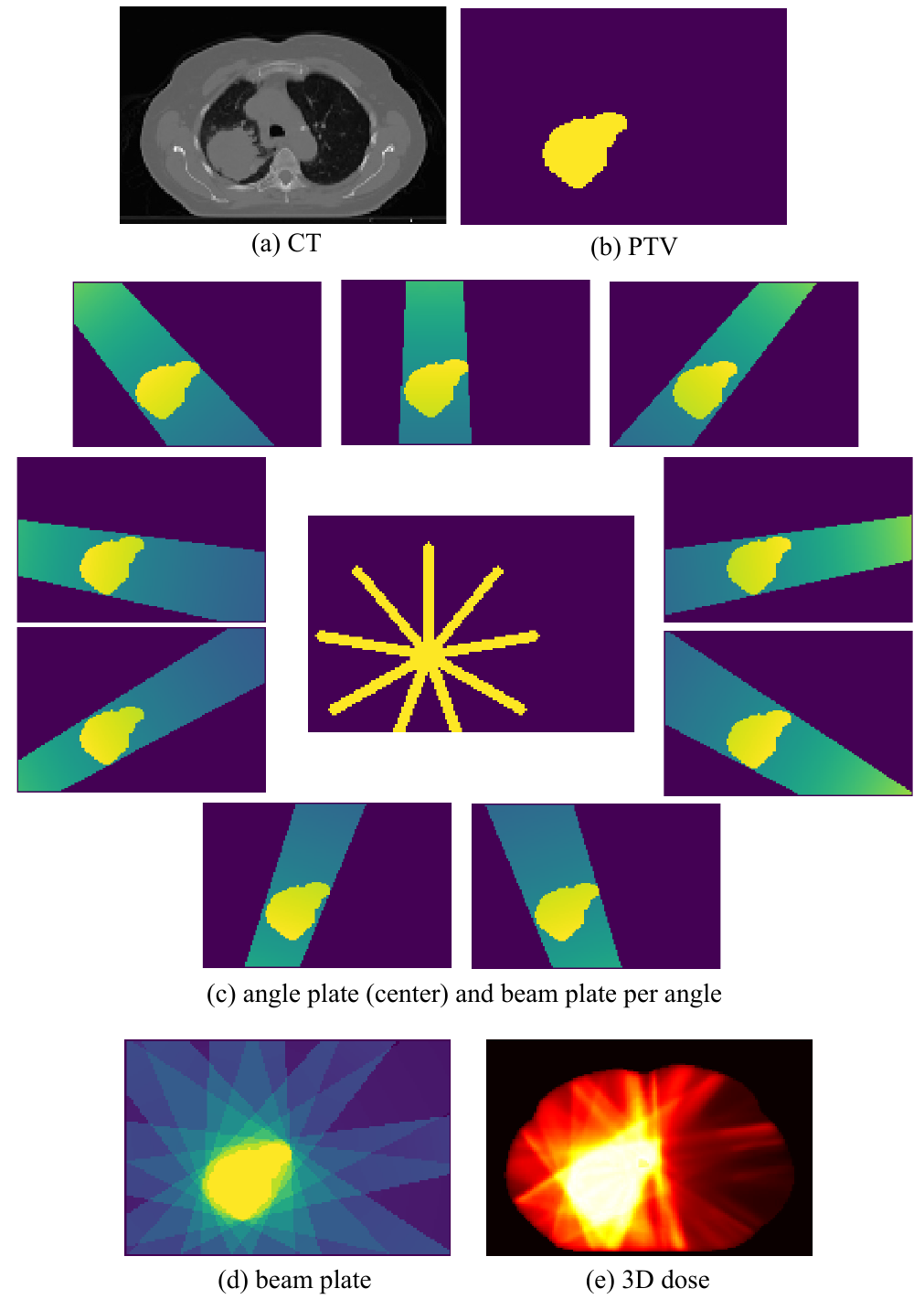}
    \caption{The example visualization of one plan with angle and beam plates. Except the angle plate, all other data are 3D though only one middle slice are used for demonstration purpose. The PTV are included in beam plate for illustrating how plates are created, but it is not necessary to be saved as part of the plate.}
    \label{fig:supp_plate}
\end{figure}

We include the creation of angle and beam plates in our \href{https://github.com/RiqiangGao/GDP-HMM_AAPMChallenge}{GDP-HMM GitHub}, which is inspired by \cite{gao2023flexible}. The angle plate gives a sketch of the angle directions, and the beam plate estimates the beam coverage considering the PTV shapes. Please follow details in the GitHub repository. 

\newpage
\section{Visualization of More Data Examples}
\label{sec:supp_moreexp}
\vspace{-0.2in}
\begin{figure}[h]
    \centering
    \includegraphics[width=0.90\linewidth]{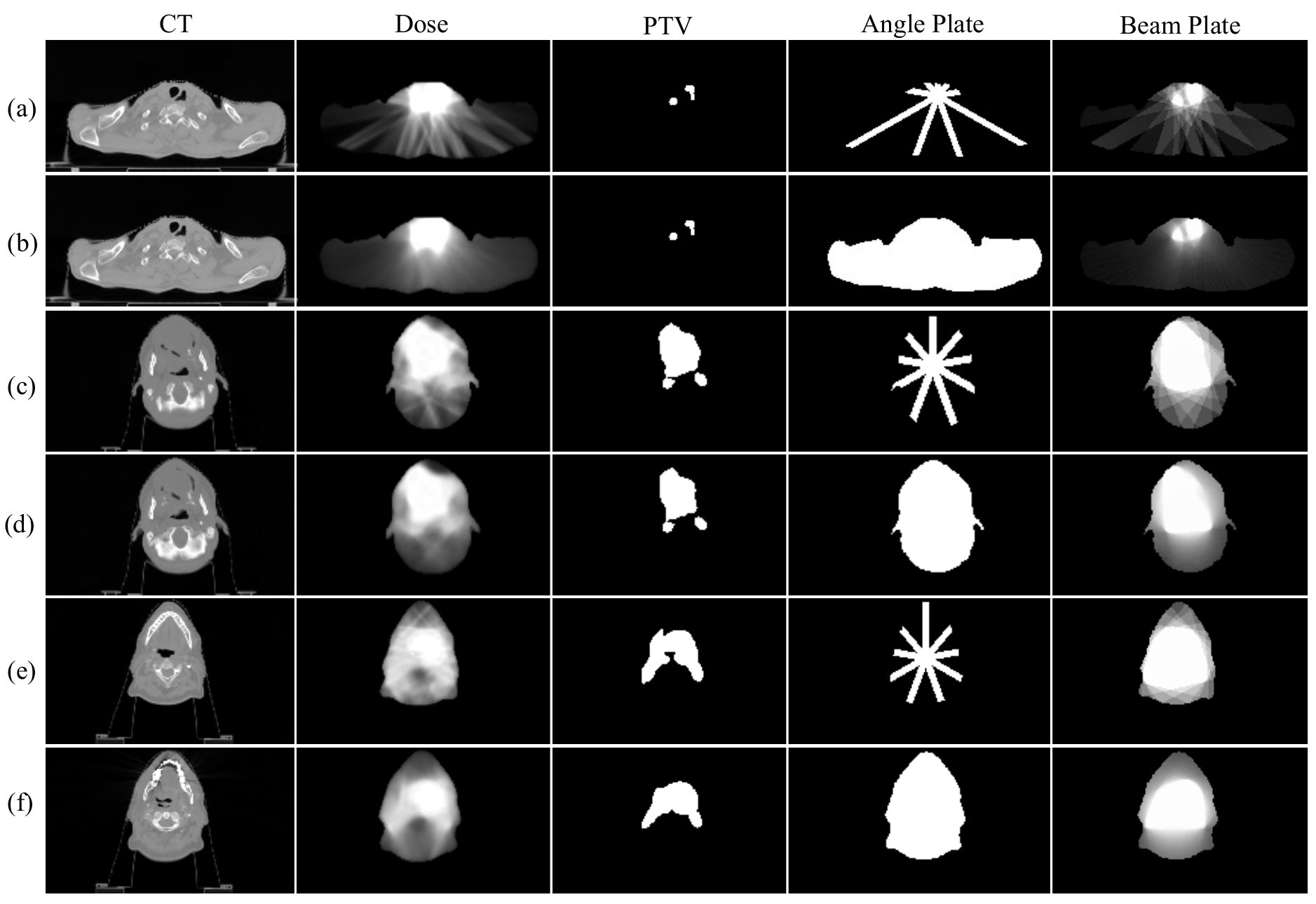}
    \vspace{-0.1in}
    \caption{Six plans from three patients for head-and-neck cancer.}
    \label{fig:supp_han_exp}
\end{figure}
\vspace{-0.2in}
\begin{figure}[h]
    \centering
    \includegraphics[width=0.90\linewidth]{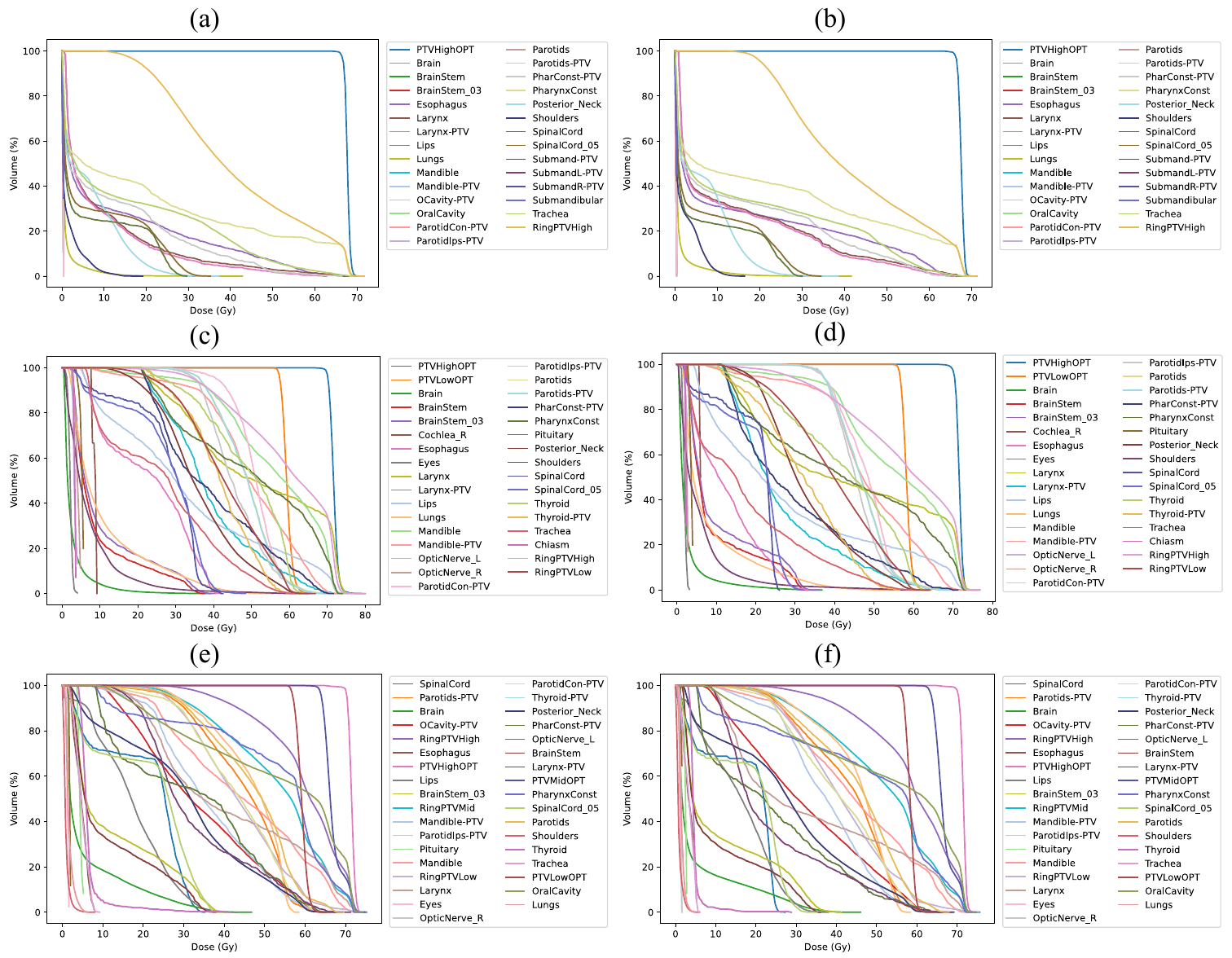}
    \vspace{-0.1in}
    \caption{The DVHs of the plans shown in Fig. \ref{fig:supp_han_exp}.}
    \label{fig:supp_han_dvh}
\end{figure}

The creation of angle and beam plates is shown in Sec. \ref{sec:supp_plate} and documented at \href{https://github.com/RiqiangGao/GDP-HMM_AAPMChallenge}{GDP-HMM GitHub}. The selected three patients have one, two and three PTVs, respectively. From the DVHs, we could visually validate the quality of our plans. 

\vspace{-0.1in}
\begin{figure}
    \centering
    \includegraphics[width=0.90\linewidth]{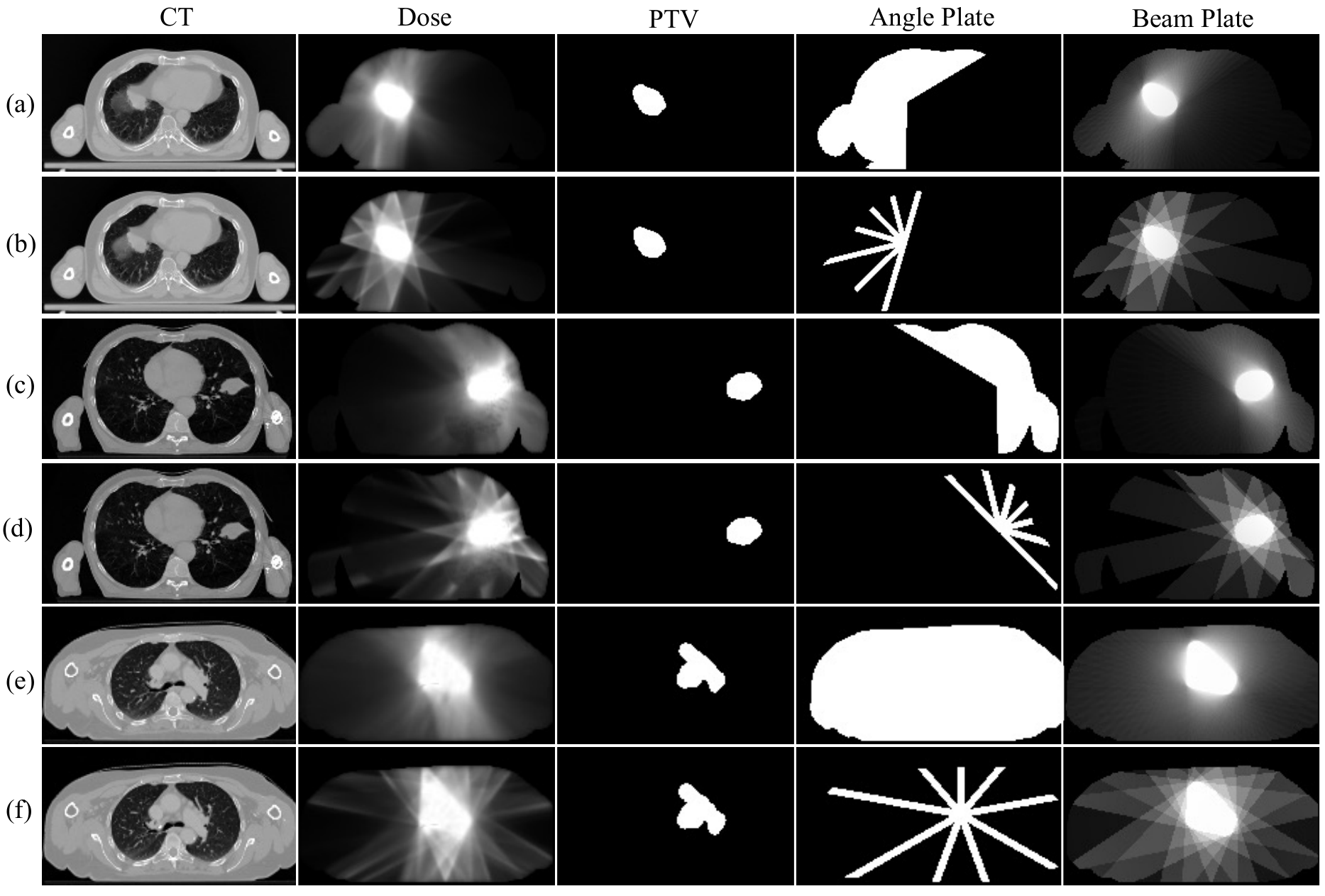}
    \vspace{-0.1in}
    \caption{Six plans from three patients for lung cancer.}
    \label{fig:supp_lung_exp}
\end{figure}
\vspace{-0.2in}
\begin{figure}
    \centering
    \includegraphics[width=0.90\linewidth]{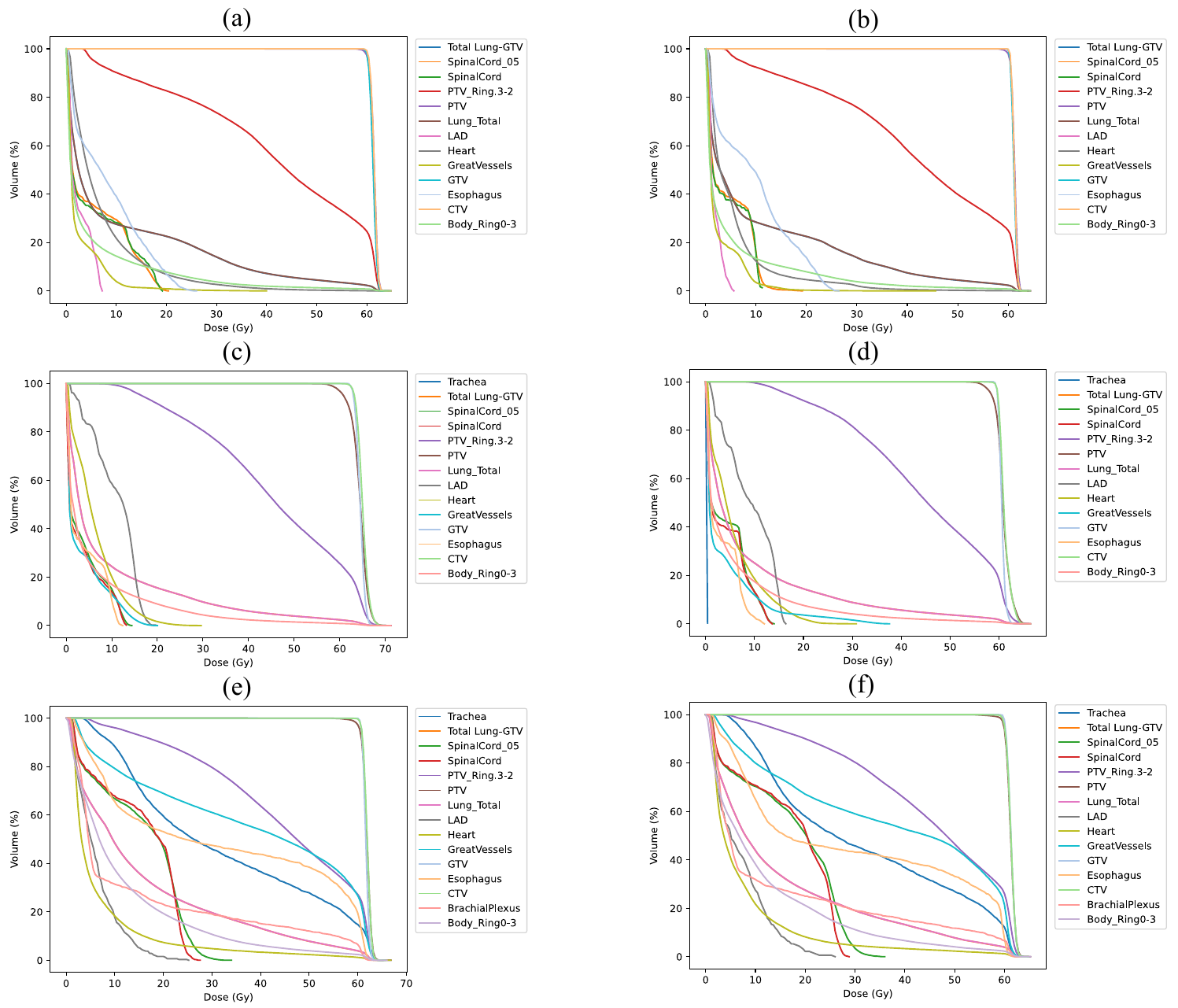}
    \vspace{-0.1in}
    \caption{The DVHs of the plans shown in Fig. \ref{fig:supp_lung_exp}.}
    \label{fig:supp_lung_dvh}
\end{figure}

\end{document}